%% file: mnras_template.tex
\documentclass[fleqn,usenatbib]{mnras}

\usepackage{mathptmx}

\usepackage[T1]{fontenc}

\DeclareRobustCommand{\VAN}[3]{#2}
\let\VANthebibliography\thebibliography
\def\thebibliography{\DeclareRobustCommand{\VAN}[3]{##3}\VANthebibliography}

\usepackage{graphicx}	

\usepackage{multicol}        
\usepackage{bm}		
\usepackage{pdflscape}	
\DeclareGraphicsExtensions{.pdf,.png,.jpg,.jpeg,.eps}
\graphicspath{{./Figures/}}
\usepackage{acronym}
\usepackage{xcolor}
\usepackage[most]{tcolorbox}
\usepackage{ulem}

\newcommand{\be}{\begin{equation}}
\newcommand{\ee}{\end{equation}}
\newcommand{\bea}{\begin{eqnarray}}
\newcommand{\eea}{\end{eqnarray}}
\newcommand{\bel}{\begin{align}}
\newcommand{\eel}{\end{align}}

\def\gccm{{\rm g\,cm^{-3}}}
\def\Msun{{\rm M_{\odot}}}
\def\GMc2{{\rm G M_{\odot} c^{-2}}}

\def\kt2{\kappa^\text{T}_2}

\newcommand{\epol}{\mathcal{E}_{pol}}
\newcommand{\etor}{\mathcal{E}_{tor}}
\newcommand{\emag}{\mathcal{E}_{\rm mag}}

\newcommand{\avgB}{\langle B \rangle}

\usepackage{pifont} 

\newcommand{\alfven}{\ensuremath{\mbox{Alfv\'{e}n}}}
\newcommand{\sm}[2]{{\tt {#1}{#2}}}

\usepackage{color}
\definecolor{cyan}{rgb}{0,0.9,0.9}
\definecolor{orange}{rgb}{0.9,0.5,0}
\definecolor{magenta}{rgb}{1,0,1}
\definecolor{purple}{rgb}{0.8,0.4,0.8}
\definecolor{gray}{rgb}{0.8242,0.8242,0.8242}

\graphicspath{{./Figures/}}

\title[GRMHD simulations of magnetic field in neutron stars]{Long-term {GRMHD simulations} of magnetic field in isolated neutron stars}

\author[A. Sur et al.]{
Ankan Sur,$^{1}$\thanks{E-mail: ankansur@camk.edu.pl}
William Cook,$^{2}$
David Radice,$^{3,4,5}$
Brynmor Haskell,$^{1}$
Sebastiano Bernuzzi$^{2}$
\\
$^{1}$Nicolaus Copernicus Astronomical Center, Polish Academy of Sciences, Bartycka 18, 00-716, Warsaw, Poland\\
$^{2}$Theoretisch-Physikalisches Institut, Friedrich-Schiller-Universit{\"a}t Jena, 07743, Jena, Germany\\
$^{3}$Institute for Gravitation \& the Cosmos, The Pennsylvania State University, University Park PA 16802, USA\\
${}^4$
Department of Physics, The Pennsylvania State University, University Park PA 16802, USA \\
${}^5$Department of Astronomy \& Astrophysics, The Pennsylvania State University, University Park PA 16802, USA 
}

\date{Accepted XXX. Received YYY; in original form ZZZ}

\pubyear{2015}

\begin{document}
\label{firstpage}
\pagerange{\pageref{firstpage}--\pageref{lastpage}}
\maketitle

\begin{abstract}
Strong magnetic fields play an important role in powering the
        emission of neutron stars. Nevertheless, a full understanding of the interior configuration of the field remains elusive. In this work, we present General Relativistic
MagnetoHydroDynamics simulations of the magnetic field evolution in
neutron stars lasting {${\sim} {880}\,$ms (${\sim} {6.5}$ \alfven \,crossing periods) and up to resolutions of $0.1155\,$km} using {\tt Athena++}.
We explore two different initial conditions, one with purely poloidal magnetic field and
the other with a dominant toroidal component, and study the poloidal and toroidal field energies,
the growth times of the various instability-driven oscillation modes and turbulence.
We find that the purely poloidal setup generates a toroidal field
which later decays exponentially reaching $1\%$ of the total magnetic
energy, showing no evidence of reaching
equilibrium. The initially stronger toroidal field setup, on the other
hand, loses up to {20}\% of toroidal energy and maintains
this state till the end of our simulation. We also explore the
hypothesis, drawn from previous MHD simulations, that turbulence plays
an important role in the quasi equilibrium state. An analysis of the
spectra in our higher resolution setups reveal, however, that in most
cases we are not observing turbulence at small scales, but rather a
noisy velocity field inside the star. 
We also observe that the majority of the magnetic energy gets dissipated as heat increasing the internal energy of the star, while a small fraction gets radiated away as electromagnetic radiation.
\end{abstract}




\section{Introduction}
\input{introduction}

\section{Method}
\label{method}
\input{method}

\section{Results}
\label{results}
\input{results}


\section{Conclusions and Discussions}
\label{conclusions}
\input{conclusions}


\section*{Acknowledgements}

	A.S thanks Lorenzo Gavassino for the useful discussion on computing energies in general relativity.
  A.S and B.H. were supported by the National
  Science Centre, Poland (NCN), via an OPUS grant number 2018/29/B/ST9/02013 and a
  SONATA BIS grant number 2015/18/E/ST9/00577.
  D.R. acknowledges support from the U.S. Department of Energy, Office of Science, Division of Nuclear Physics under Award Number(s) DE-SC0021177 and from the National Science Foundation under Grants No. PHY-2011725 and PHY-2116686.
  S.~B. acknowledges support by the EU H2020 under ERC Starting
	Grant, no.~BinGraSp-714626.  

	Computations were performed on the ARA cluster at Friedrich
	Schiller University Jena, on the supercomputer SuperMUC-NG at the
	Leibniz-Rechenzentrum (LRZ, \url{www.lrz.de}) Munich, and on the
	national HPE Apollo Hawk at the High Performance Computing
	Center Stuttgart (HLRS).
	The ARA cluster is funded in part by DFG grants INST
	275/334-1 FUGG and INST 275/363-1 FUGG, and ERC Starting Grant, grant
	agreement no. BinGraSp-714626.
	The authors acknowledge the Gauss Centre for Supercomputing
	e.V. (\url{www.gauss-centre.eu}) for funding this project by providing
	computing time to the GCS Supercomputer SuperMUC-NG at LRZ
	(allocations {\tt pn56zo}, {\tt pn68wi}).
	The authors acknowledge HLRS for funding this project by providing
	access to the supercomputer HPE Apollo Hawk under the grant
	number {\tt INTRHYGUE/44215}.

\section*{Data Availability}

The data underlying this article would be available on request to the corresponding author.



\bibliographystyle{mnras}
\bibliography{references} 



\appendix
\section{}
\input{appendix}
\label{app}


\bsp	
\label{lastpage}
\end{document}

%% file: introduction.tex
Harboring the strongest magnetic fields in the universe with core densities exceeding that of nuclear matter, neutron stars (NSs) provide a laboratory for studying physics at extreme conditions, which are not reproducible with the current available technologies on {Earth}. The surface magnetic field ($B_s$) of NSs is generally inferred from the dipole spin-down using radio astronomical data \citep{Chung1,Chung2} and have allowed us to classify these systems into old-recycled pulsars with $B_s \sim 10^{8}$ G, ordinary pulsars with $B_s \sim 10^{12}$ G and magnetars with $B_s \sim 10^{15}$ G. Alternatively, the geometry of the magnetic field and its strength had been derived from X-ray emitting hotspots in pulsar PSR-J001X \citep{Bilous2019}. This study (see also \citep{deLima2020}) suggests that the field is far from the conventional dipolar geometry, but rather favors a multipolar magnetic field or an offset-dipole.
	
The magnetic field of pulsars plays an important role in accelerating charged particles in the magnetospheres which emit electromagnetic radiation and allows us to study its properties, for example spin down due to magnetic dipole radiation. The energy from differential rotation can be converted to a large-scale magnetic field which in turn can help in launching powerful jets {from newly formed NSs} \citep{Moiseenko2006,Shibata2006,Burrows2007,Mosta2014}. It has been observed that this magnetic field remains stable on a longer timescale comparable to the lifetime of NSs except that of magnetic flares emitted by magnetars, which operate on a very short timescale. This provokes the quest to understand what leads to the magnetic field stability. Although there is evidence of an exponential decay of the field through Ohmic dissipation, the timescale responsible for this mechanism \citep{Ostriker1969} is greater than the Hubble time. In other words, there is no significant decay that changes the magnetic field effectively \citep{Krav2021}.

An arbitrary magnetic field is generally not in equilibrium when the Lorentz force and pressure forces do not balance one another. It had been long established that certain equilibrium configurations, like a purely poloidal or a purely toroidal field, is unstable and subjected to ``kink'' instability acting within few Alfven timescales \citep{Tayler57,Tayler73,Wright73,Markey73,Markey74,Flowers77}. Analytical \citep{Haskell2008, Ciolfi2009, Ciolfi2010, Gusakov2017, Gusakov2018} and numerical simulations including Newtonian magnetohydrodynamics (MHD) \citep{BraithwaiteSpruit2005,BraithwaiteNord2005,Braithwaite2007,Lander2009,Lander2010,Lander2011,Herbrik2017,Frederick2020, Sur2020} and General Relativistic (GR) MHD \citep{Kiuchi2008,Ciolfi2011,Ciolfi2013,Lasky2011,Pili14,Pili17} have confirmed this explicitly where a stable stellar field needs both poloidal and toroidal components. 
The instability gives rise to various azimuthal oscillation modes responsible for driving gravitational radiation from the system. Estimating the relative strength of the poloidal and toroidal components is also important in virtue of studying continuous gravitational waves emitted by NSs, caused by magnetic deformation \citep{Bonazzola1996,Cutler2002,Frieben2012}, as it depends sensitively on the amount of magnetic field energy stored in each of its components. 
		
The most favored {magnetic field} geometry pertinent to NSs is that of a ``twisted-torus'' where the poloidal field lines thread the interior of the star and close inside. Outside the star, the field lines extend {until} infinity with the field being continuous at the stellar surface. This implies the absence of surface currents. The toroidal field is concentrated as a flux tube within this closed poloidal field line located at approximately 0.8 times the stellar radii. This geometry had been {found} by time-evolving random initial configurations from MHD  simulations \citep{BraithwaiteNord2005}. 

{Understanding equilibria requires us to solve the so-called Grad-Shafranov equation which yields various magnetic field configurations with varying poloidal and toroidal field energies \citep{Lander2009,Ciolfi2012, Ciolfi2013, Gourgouliatos2013,Armaza2015,Sur2021}. However, these solutions do not tell us anything about the stability of the magnetic field with time. Studies with an axisymmetric field in the crust have been performed to show the presence of Hall equilibrium states \citep{Hollerbach2002,Cumming2004,Gourgouliatos2014,KGAC2014}. The Hall effect leads to the formation of small-scale magnetic features which dissipates to power the thermal radiation, e.g. in magnetars, when a toroidal magnetic field of strength $10^{16}$ G is present inside the crust \citep{KG2016,Geppert2014,Pons2011}. The Hall effect also leads to the growth of dipole moment of a quadrupolar toroidal component in NS crust which could explain the observed braking indices of young pulsars \citep{KG2015}. Further, long-term evolution of the magnetic field in the crust of neutron stars under the Hall effect and Ohmic dissipation has shown the presence of a ``Hall attractor'' state which for an initially dipole dominated field has also octupolar component and an energetically negligible quadrupole toroidal field.}

{Due to the lack of} direct observational evidence of the internal magnetic field topology, our knowledge on the distribution of magnetic energy in the poloidal and toroidal components is limited to simulations. It had been shown that about $80-90\%$ of the total magnetic energy is stored within the poloidal component {from MHD simulations with either a purely initial poloidal field or from a mixed-field with a stronger toroidal component \citep{Sur2020}. However, equilibria calculations by \cite{Ciolfi2013} have shown to produce toroidal field energy 90\% of the total internal magnetic energy with a suitable choice of the azimuthal currents. Whether these models are realistic demands future studies not only to validate their stability but also for a better understanding of emission properties from NSs}. Another interesting finding by \cite{Sur2020} is that NSs experience turbulence triggered by the initial perturbations to the field. It had been shown that this turbulence gives rise to an inverse cascade in magnetic helicity ($H_m$) which determines the ``twist'' of the magnetic field lines. Thus energy is transferred from small resistive scales to large eddies. Further, the conservation of $H_m$ is broken as the field rearranges and attends stability. It was found that the energy spectra followed Kolmogorov law with a scaling $-5/3$, but the data were noisy owing to limited resolution using the spherical coordinate system. {These simulations studied the first 40 ms of magnetic field evolution. We try to understand turbulence at late times $t>100$ ms for which we need longer simulations with higher resolution. Moreover, the magnetic field geometry can be used as various background models in other studies like post-merger BNS simulations in which the magnetic field is either responsible for a strong baryonic wind \citep{Kalinani2020} or responsible for jet formation, powering kilonova transients and GW emission \citep{Ciolfi2020}.}

In this study, we perform nonlinear GRMHD simulations for a fiducial NS of mass $1.4 \,{M_{\odot}}$ using the code {\tt Athena++} as described in section \ref{method}. The results of the simulations are presented in section \ref{results} which are some of the longest in terms of evolution time and hence gives us further insights on what happens to the magnetic field energy and its structure at later times. We also investigate the relativistic effects of turbulence and energy cascades and seek answers to the question on whether the turbulent feature persists in NS simulations. Finally, the conclusions and discussions are presented in section \ref{conclusions}.

%% file: method.tex
\begin{figure*}
	\centering
	\includegraphics[width=0.33\linewidth]{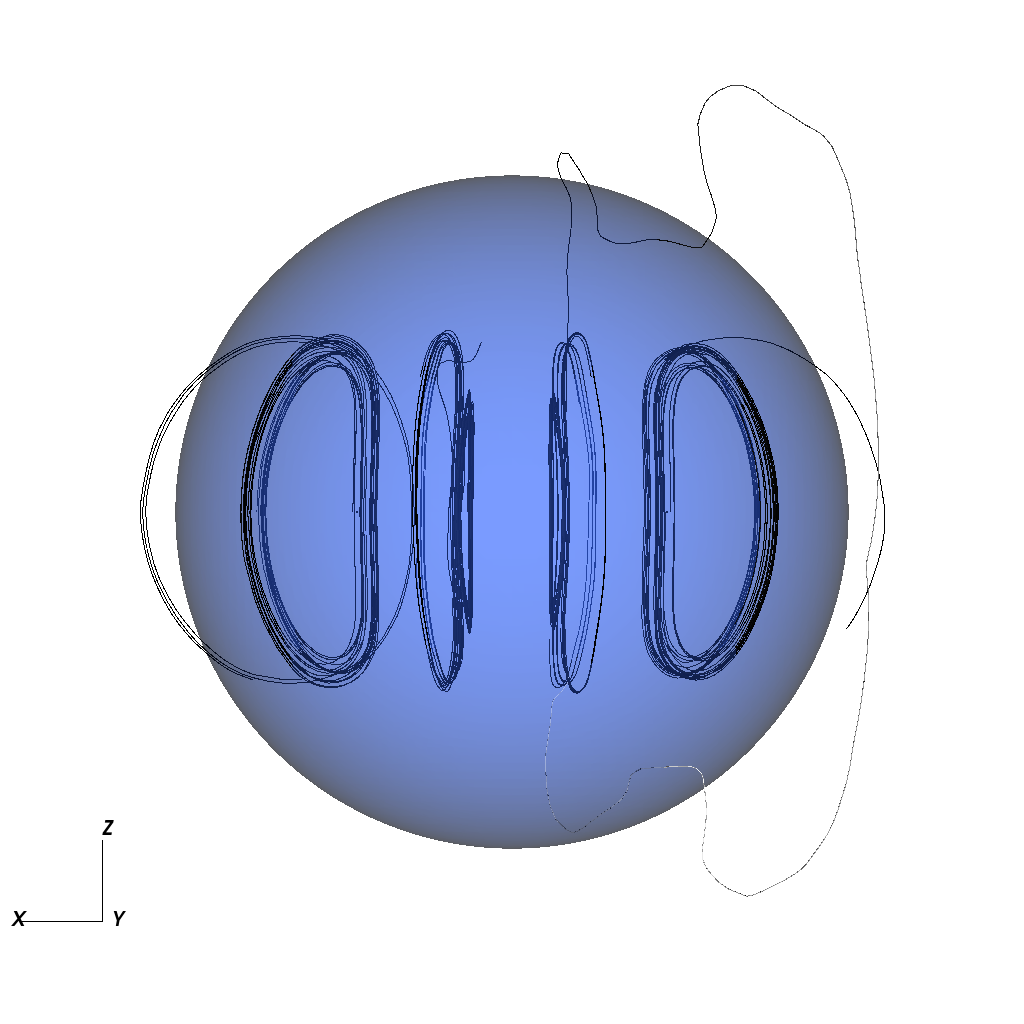}
	\includegraphics[width=0.33\linewidth]{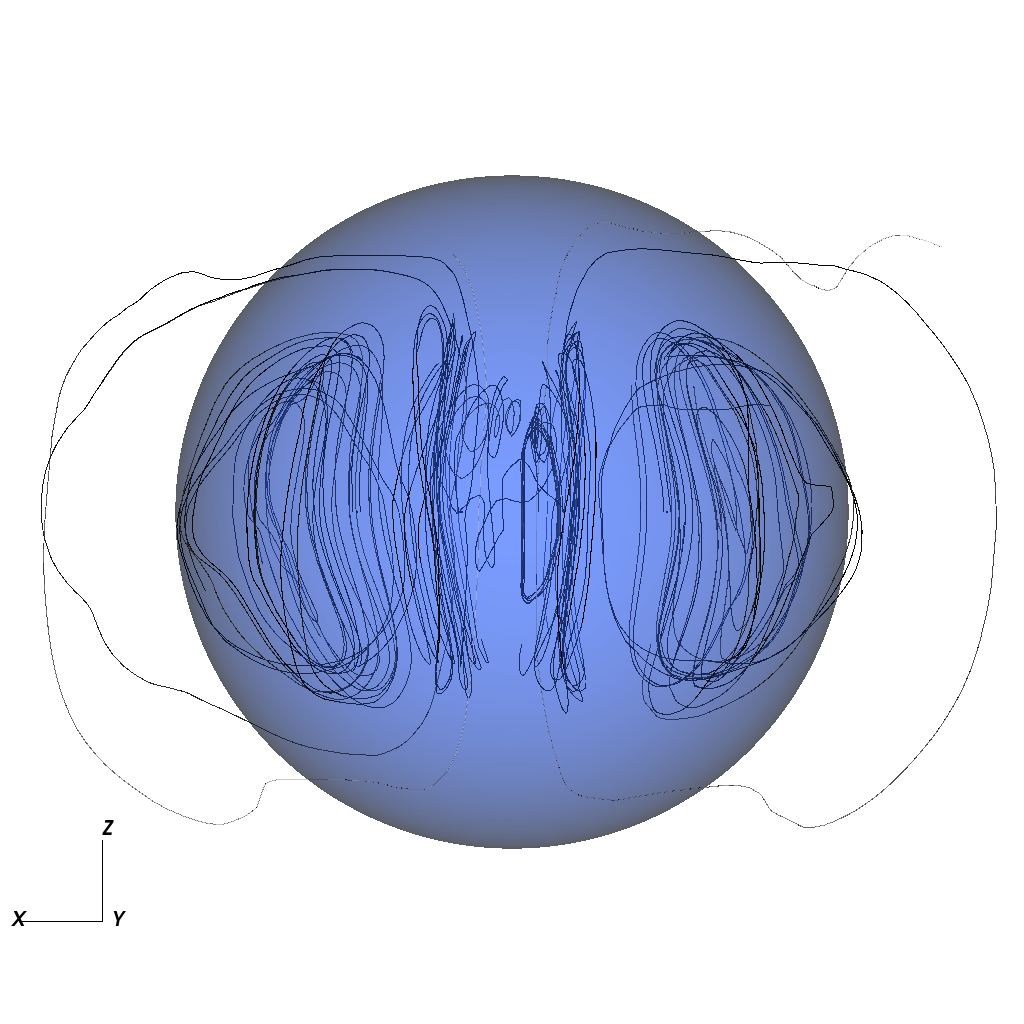}
	\includegraphics[width=0.33\linewidth]{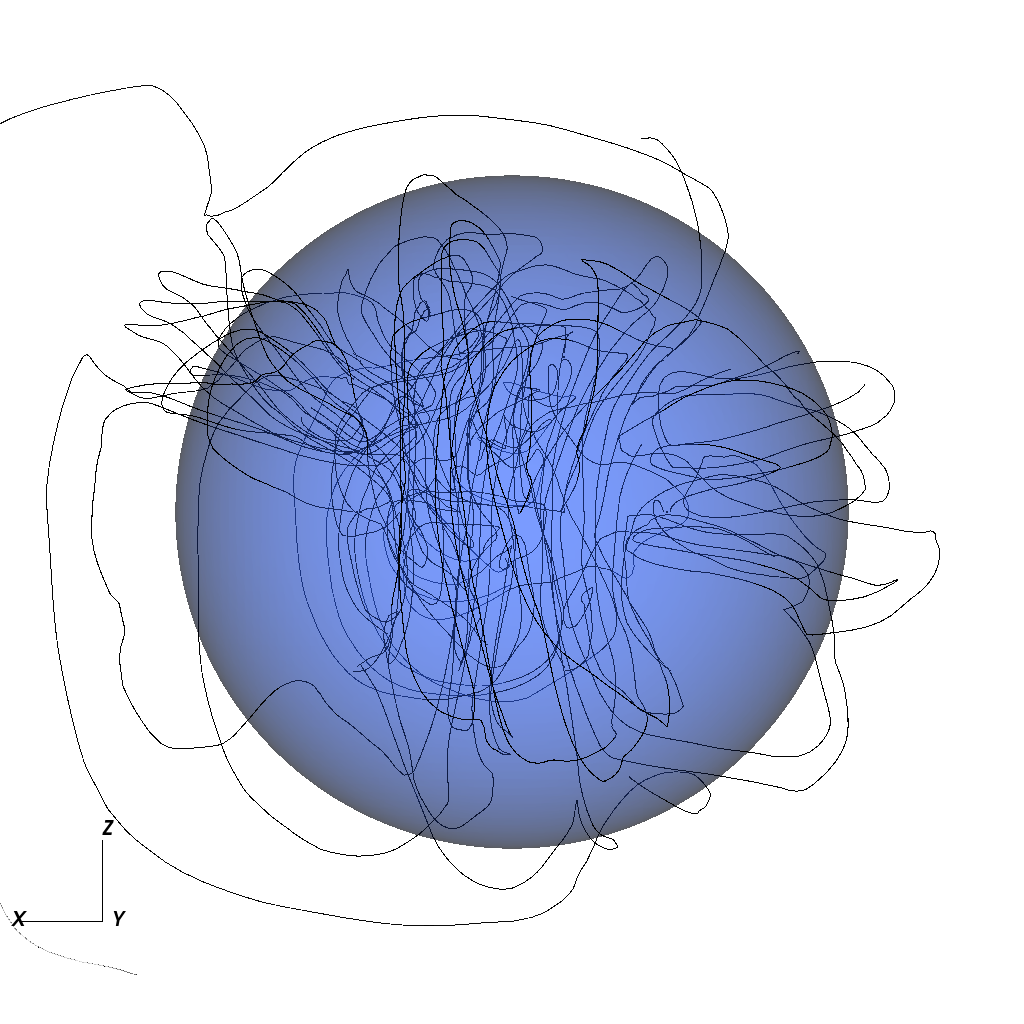}
	\includegraphics[width=0.33\linewidth]{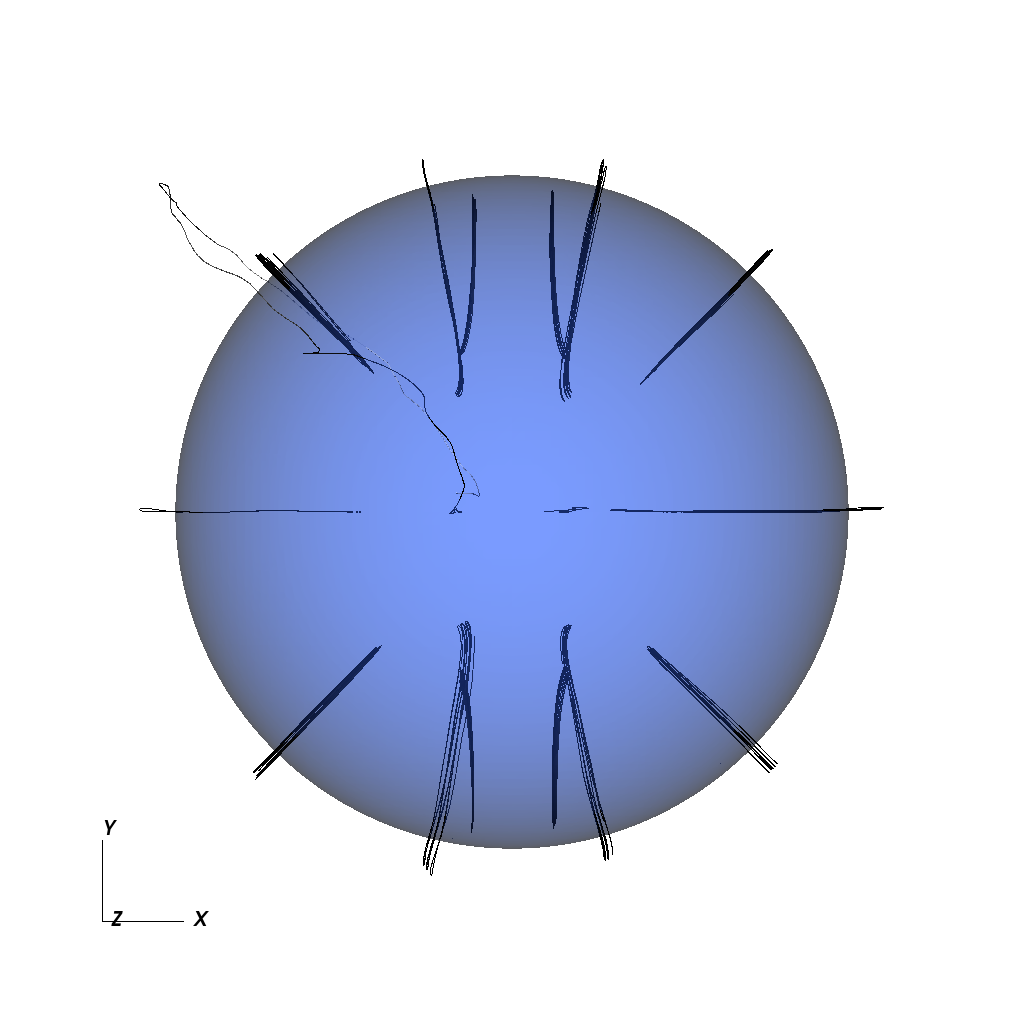}
	\includegraphics[width=0.33\linewidth]{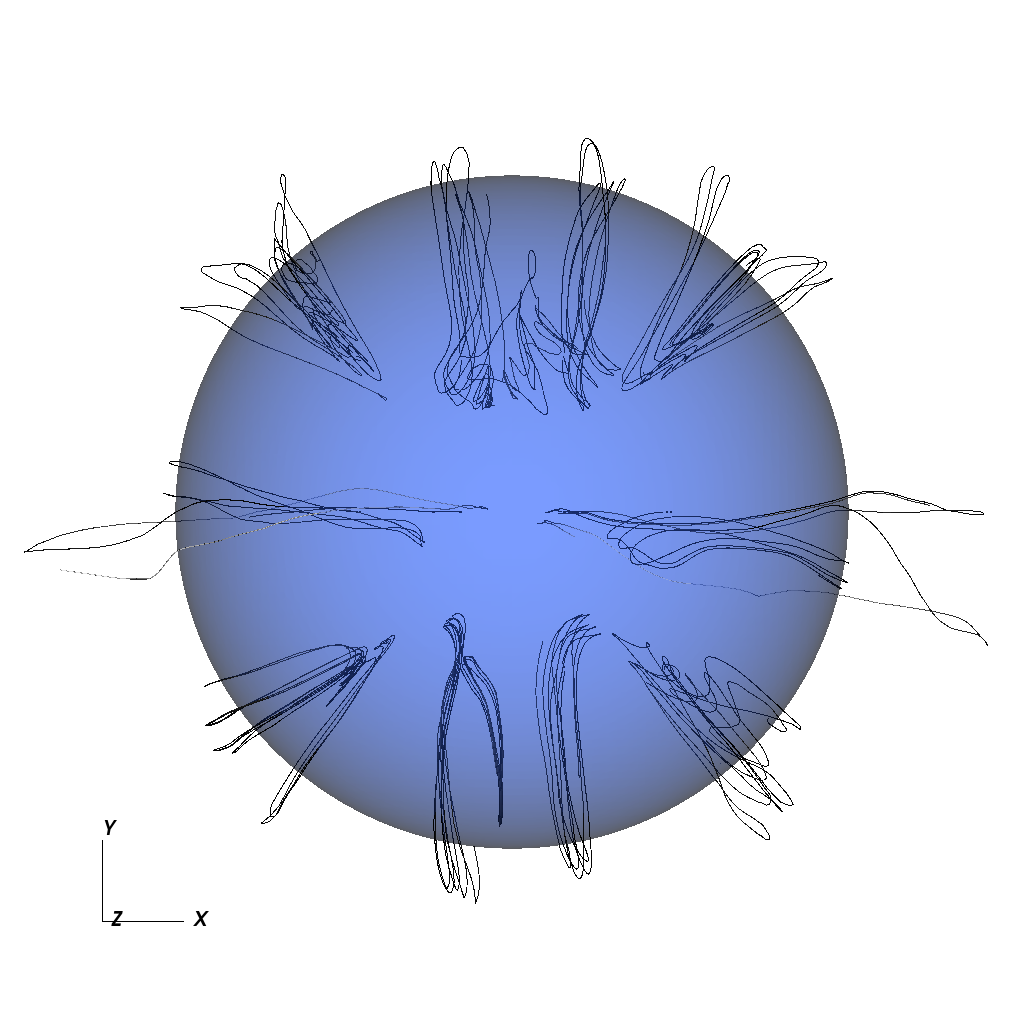}
	\includegraphics[width=0.33\linewidth]{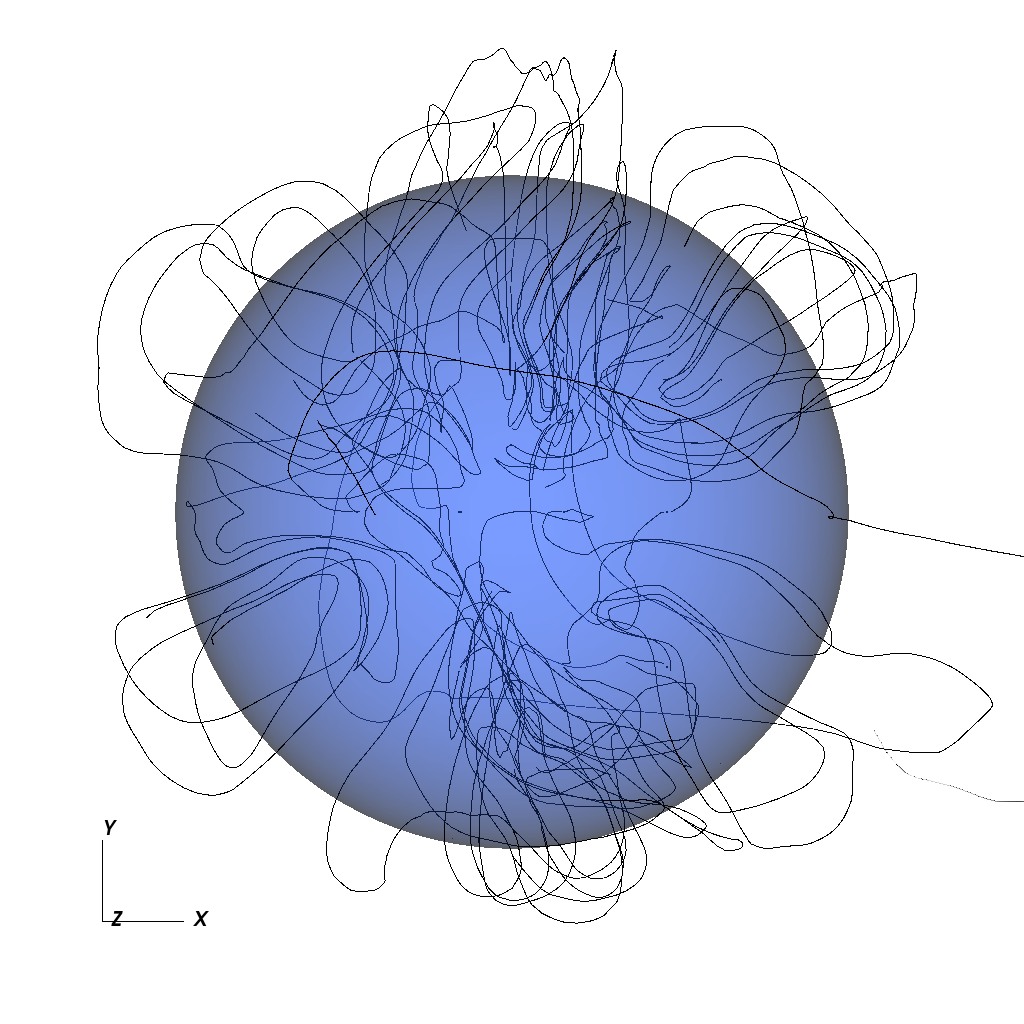}
	\caption{Three dimensional view of the magnetic field {for \sm{ps}{256}} at three different times, {$t=2$ ms ($T_A \sim 0.1$), $t=12.7$ ms ($T_A \sim 0.8$), $t=49$ ms ($T_A \sim 2$). The definition of $T_A$ is given by equation \ref{alfperiod}}. The top row shows from meridional view while the bottom row shows the equatorial view.}
	\label{fig3d}
\end{figure*}

To study the evolution of the magnetized star we perform numerical simulations using the GRMHD code {\tt Athena++} \cite{Stone_2020}. We evolve the GRMHD equations without resistivity on a fixed background metric (i.e.\ in the Cowling approximation) of a non-rotating star. {\tt Athena++} {uses} a constrained transport algorithm to evolve the magnetic field, detailed in \cite{White_2016}. Evolutions are performed using the Local-Lax-Friedrichs (LLF) flux, with reconstruction performed in the primitive variables following an implementation of the Piecewise Parabolic Method (PPM) detailed in \cite{Felker_2018}. During the simulation, primitive variables are recovered from the conservative variables by implementing the conservative to primitive inversion algorithm described in \cite{Noble_2006}.

Initial data for the geometry and matter is constructed by the numerical solution of the Tollmann-Oppenheimer-Volkoff (TOV)  equations for a spherically symmetric fluid distribution for the interior of the star coupled with an equation of state (EOS) in the form $p(\rho)$ that connects the pressure to the rest-mass density. The exterior set to the Schwarzschild metric. The fiducial NS has mass $1.4 \, {M_{\odot}}$ and a radius of $R\sim {10}\,$km. The {initial data} EOS is set to 
\begin{eqnarray}
  p = K \rho^\Gamma
\end{eqnarray}
with $K=100, \Gamma = 2$; the $\Gamma$-law EOS, 
\begin{eqnarray}
  p = (\Gamma-1)\rho\epsilon\,,
\end{eqnarray}
 where $\epsilon$ is the specific internal energy, is used during the evolution.

The initial poloidal magnetic field configuration is given by the vector potential
\begin{eqnarray}
	A_x &=& -yA_\varphi\\
	A_y &=& x A_\varphi\\
	A_z &=& 0\\
	A_\varphi &=& A_b \max(p-p_{thr},0) \\
	p_{thr} &=& 0.04p_{max}
\end{eqnarray}
following \cite{Liu_2008}, where $p_{\mathrm{max}}$ is the maximum value of the pressure within the star. 
The parameter $A_b$ controls the magnitude of the magnetic field, and is set to obtain a maximum value of $3.54\times10^{16} $G inside the star. 
The initial toroidal magnetic field is initialized directly on the magnetic field components to be
\begin{eqnarray}
	B_x &=& -\hat{y}B_{\mathrm{tor}}\max(p-p_{thr},0)\\
	B_y &=& \hat{x}B_{\mathrm{tor}}\max(p-p_{thr},0)\\
	B_z &=& 0
\end{eqnarray}

{In the case of the toroidal simulations in this paper $B_\mathrm{tor}$ is set to give the same maximum field strength as in the poloidal case. In addition, a weak poloidal field is superimposed on top of this toroidal field, with parameter $A_b$ set 50 times smaller than in the purely poloidal case.}
Evolutions are performed in unigrid without symmetry, with an outer boundary set at {30} km. {We use outflow boundary conditions which do not affect the dynamics in the interior of the star, however, it affects the dynamics at the outer edge of our simulation box. We perform runs at 4 resolutions corresponding to $64^3$, $128^3$, $256^3$, $512^3$ grid points across the computational domain. These give grid spacings of $0.923\,$ km, $0.462$ km, $0.231\,$ km, and $0.115$ km respectively. Our highest resolution run $512^3$ has an evolution time of $89$ ms whereas the setup \sm{ps}{256} has an evolution time of $880$ ms.} In the exterior of the star an atmosphere is set with {rest-mass} density $\rho_{\mathrm{atm}} = 10^{-10}\max{(\rho)}$. Any cells with density falling below a threshold value of $\rho_{\mathrm{thr}}  = 100\rho_{\mathrm{atm}}$ are identified as atmosphere and set to $\rho_{\mathrm{atm}}$, {with the fluid velocity set to 0 and the pressure fixed using the EOS. In the atmosphere the magnetic field components remain unrestricted}. We use the following nomenclature when referring to the different resolution setups. The initially purely poloidal magnetic field setup is denoted by \sm{pS}{???} where {\tt ???} represents the number of grid points. Thus, the setup \sm{pS}{256} corresponds to a purely poloidal initial field run with a numerical box of $256^3$ grid points. Similarly, the initially stronger toroidal field setup is denoted by \sm{tS}{???} where again {\tt ???} represents the number of grid points.

\begin{figure*}
	\includegraphics[scale=0.5]{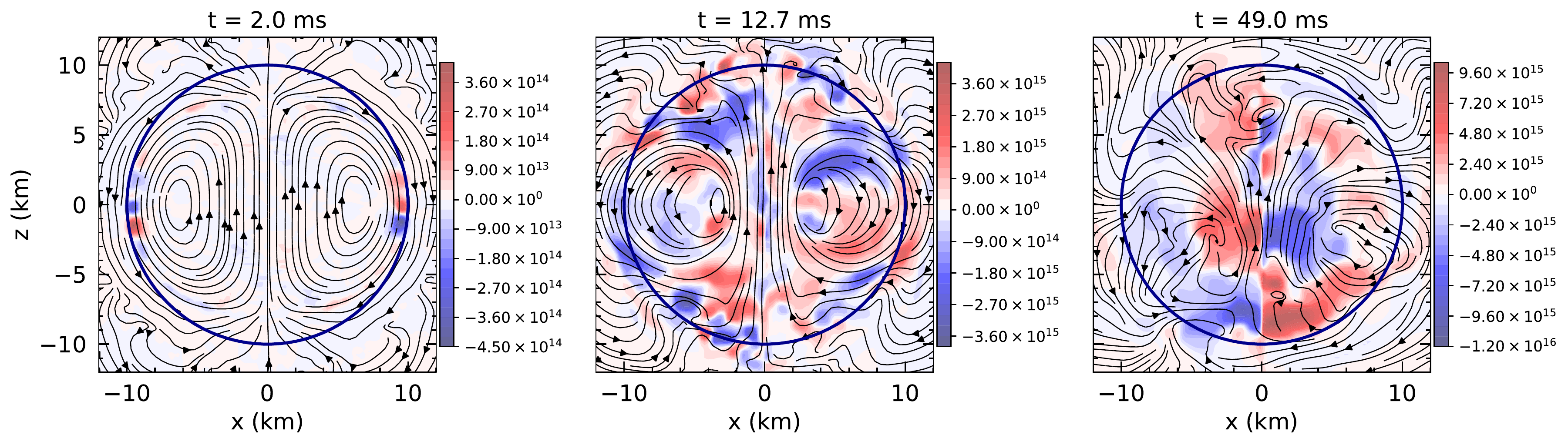}
	\includegraphics[scale=0.5]{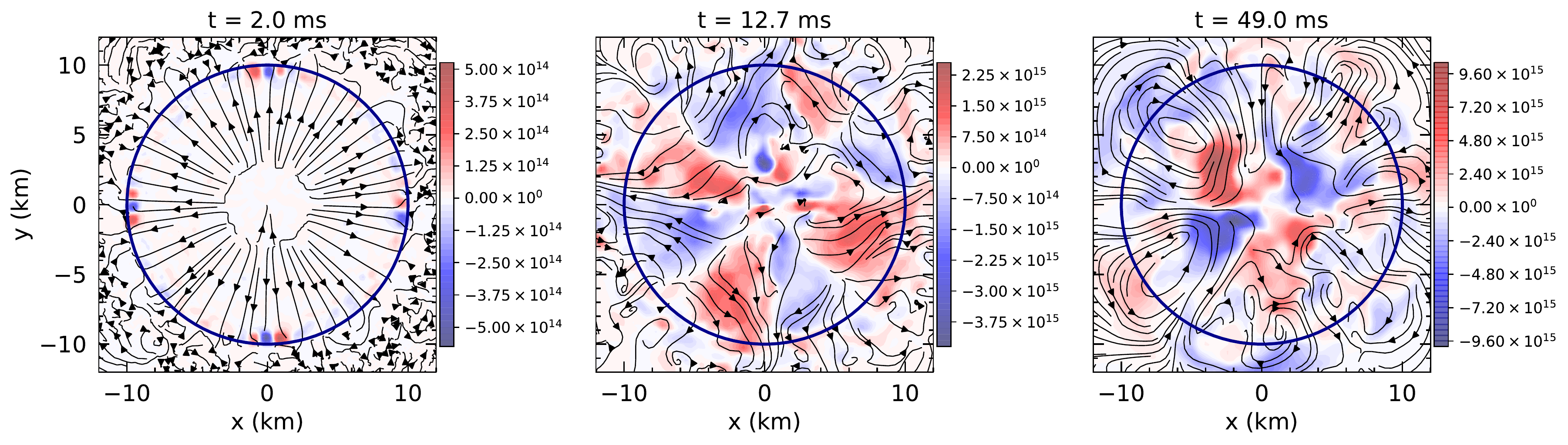}
	\caption{{Two dimensional projection of the field lines for the setup \sm{pS}{256} on the x-z plane (top row) and the x-y plane (bottom row) at the three different times, $t=2$ ms ($T_A \sim 0.1$), $t=12.7$ ms ($T_A \sim 0.8$), $t=49$ ms ($T_A \sim 2$)} as shown in Fig.~\ref{fig3d}. {The colorscale represents the strength of the toroidal field in Gauss.}}
	\label{flines}
\end{figure*}

\begin{figure*}
	\includegraphics[scale=0.45]{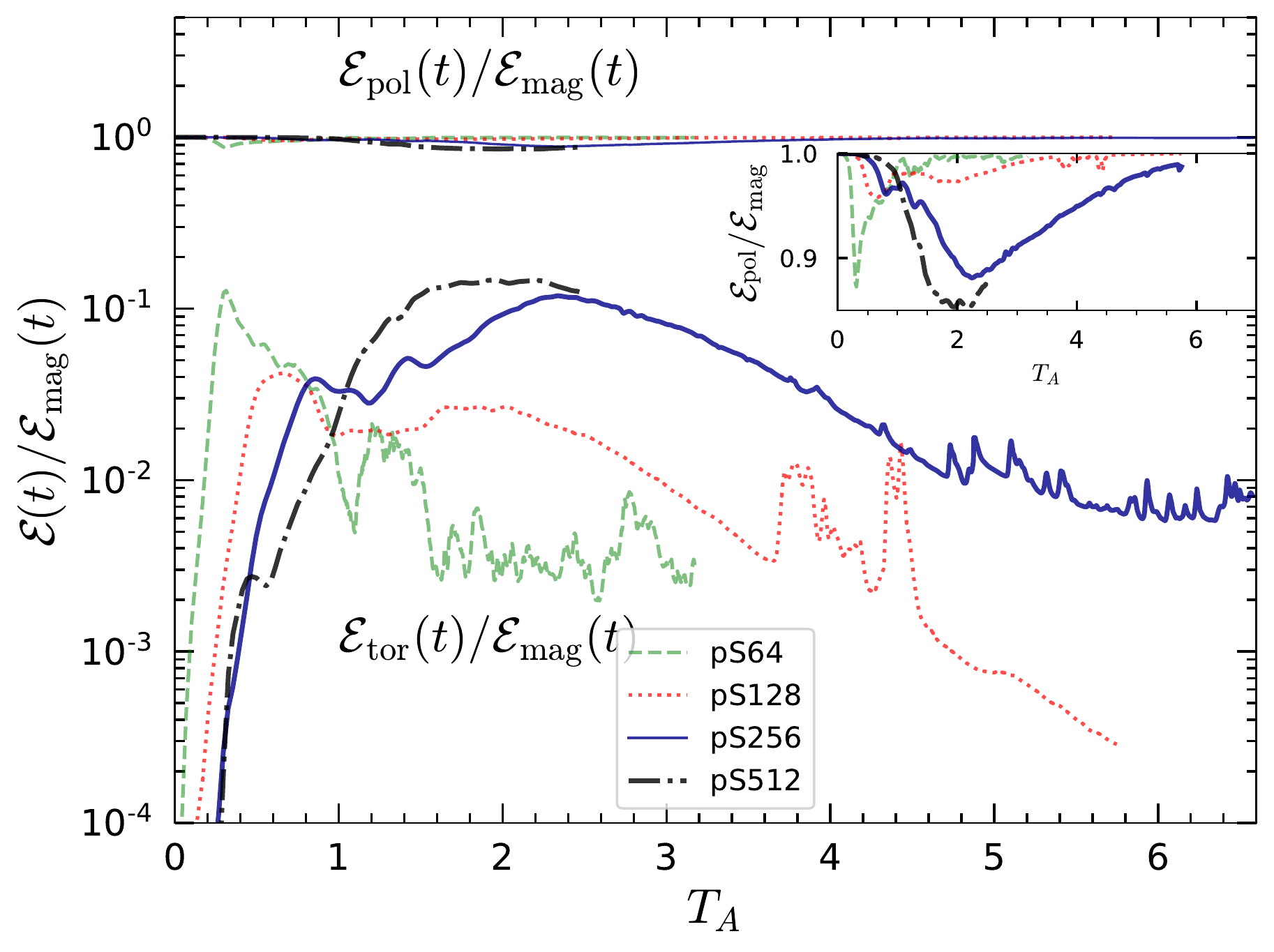}
	\includegraphics[scale=0.45]{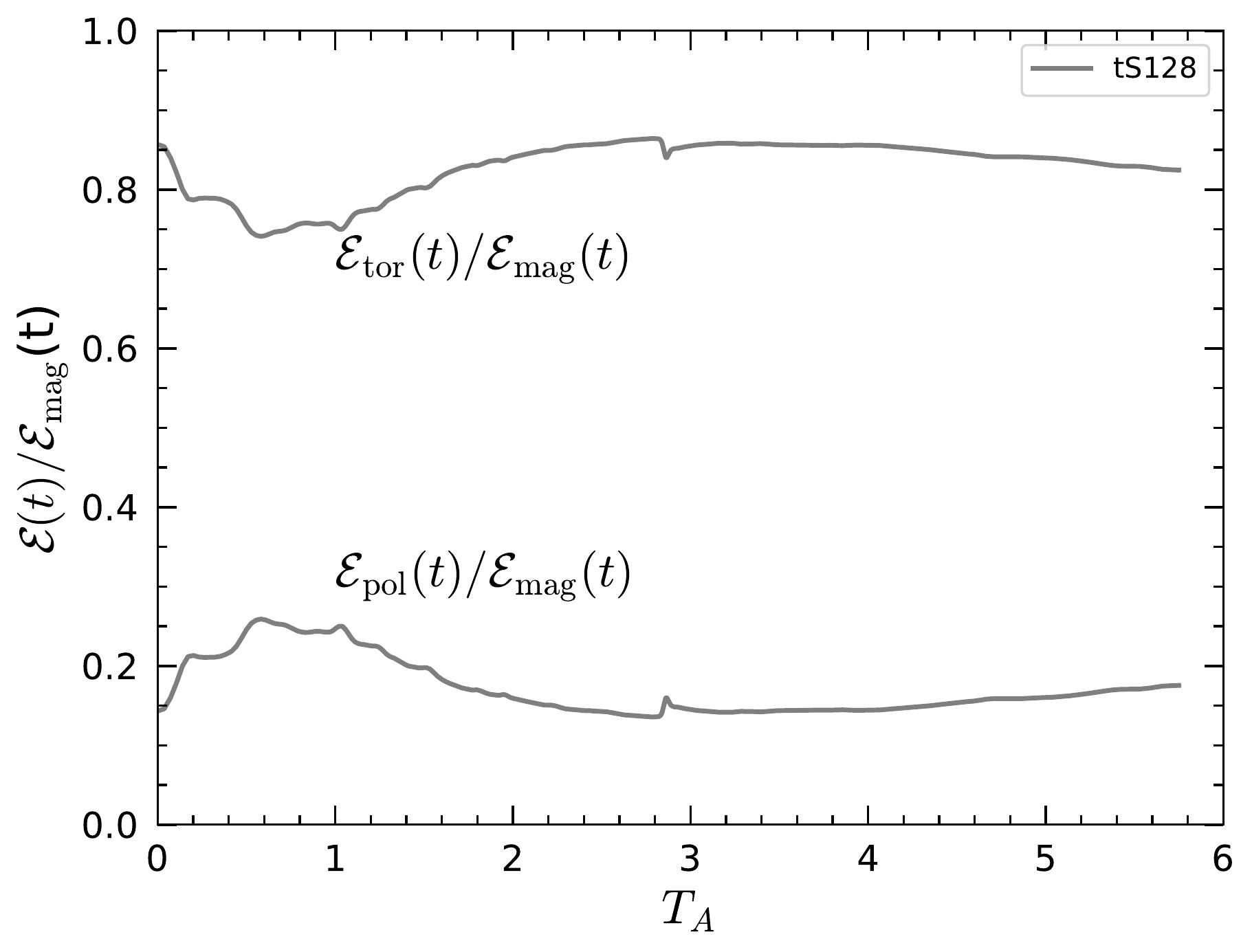}
	\caption{Ratio of poloidal and toroidal energies to the total
		magnetic energy at each \alfven \, time for (left) purely poloidally dominated initial condition (right) toroidally dominated initial condition. }
	\label{bratio_p}
\end{figure*}

%% file: results.tex
The evolution of the magnetic field occurs on a characteristic timescale associated with the system, called the $\alfven$ crossing time, which is given by
\begin{equation}
\tau_{A} = \frac{2R \sqrt{4 \pi  \langle\rho\rangle }}{\avgB}
\label{tau1}
\end{equation}
where $\langle..\rangle$ represents volume averaged quantities. For $\langle B \rangle \sim 4.5 \times 10^{15}$ G, we obtain $\tau_{A} \sim 12$ ms. Theoretically, we should expect the field to rearrange itself at this time, as we shall see it indeed does in our simulations. As $\avgB$ evolves with time, the timescale defined in Eq.~\eqref{tau1} will vary. {Because of this, we use the following definition of \alfven\, crossing period as:}
\begin{equation}
T_A = \int_0^t \frac{dt}{\tau_{A}(t)}
\label{alfperiod}
\end{equation}
{The above definition means that we are rescaling our evolution time with the \alfven \,crossing time. Hence a value of $T_A = 1$ means an evolution time equilvalent to 1 $\tau_A$, $T_A = 2$ is equivalent to 2 $\tau_A$ and so on. An evolution time of 880 ms corresponds to $\sim 6.5 \, T_A$} for our setup \sm{pS}{256}, however $T_A$ changes for the different resolution setups depending on how much magnetic energy is lost and how the density changes with time. 

\begin{figure*}
	\includegraphics[scale=0.35]{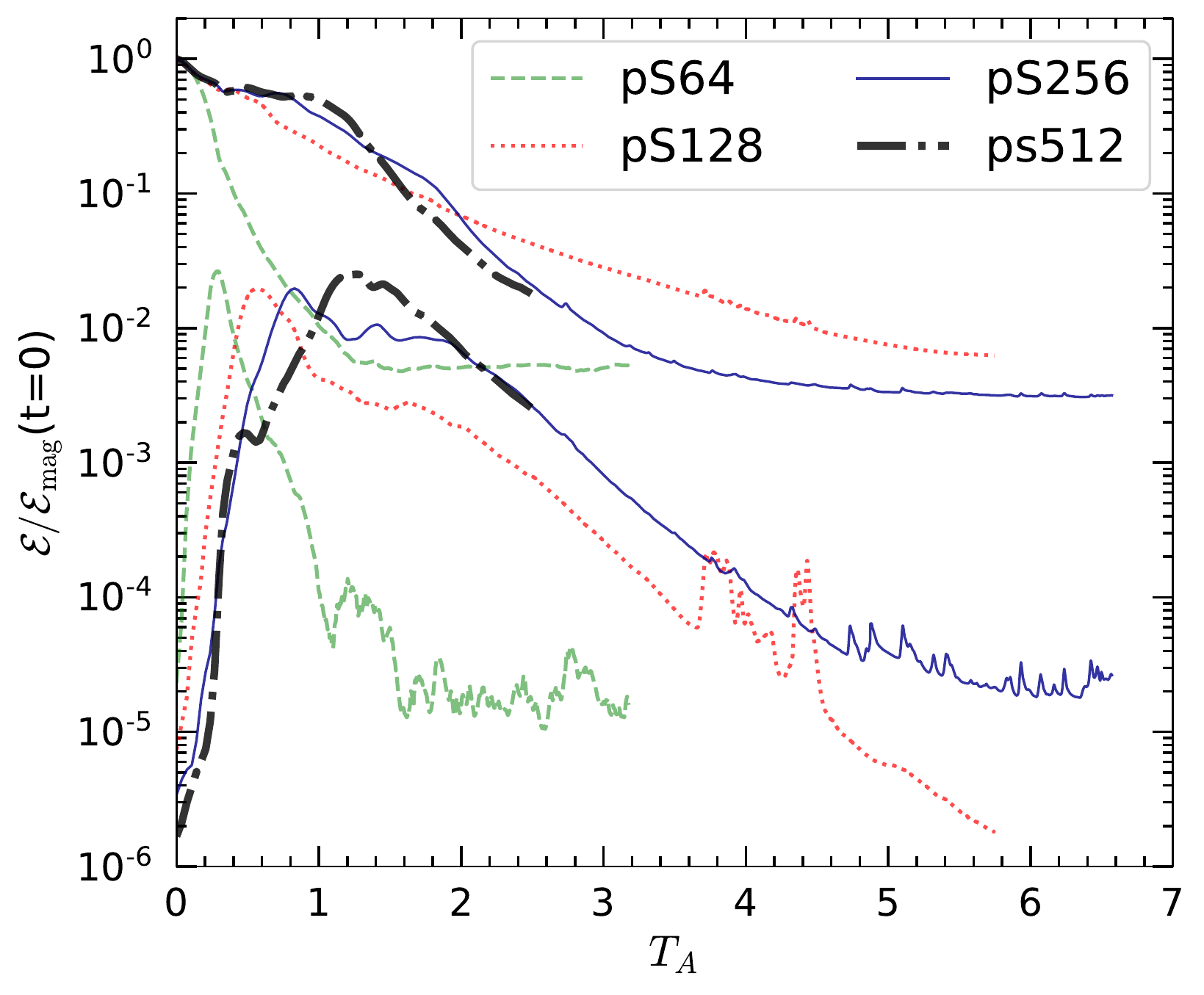}
	\includegraphics[scale=0.35]{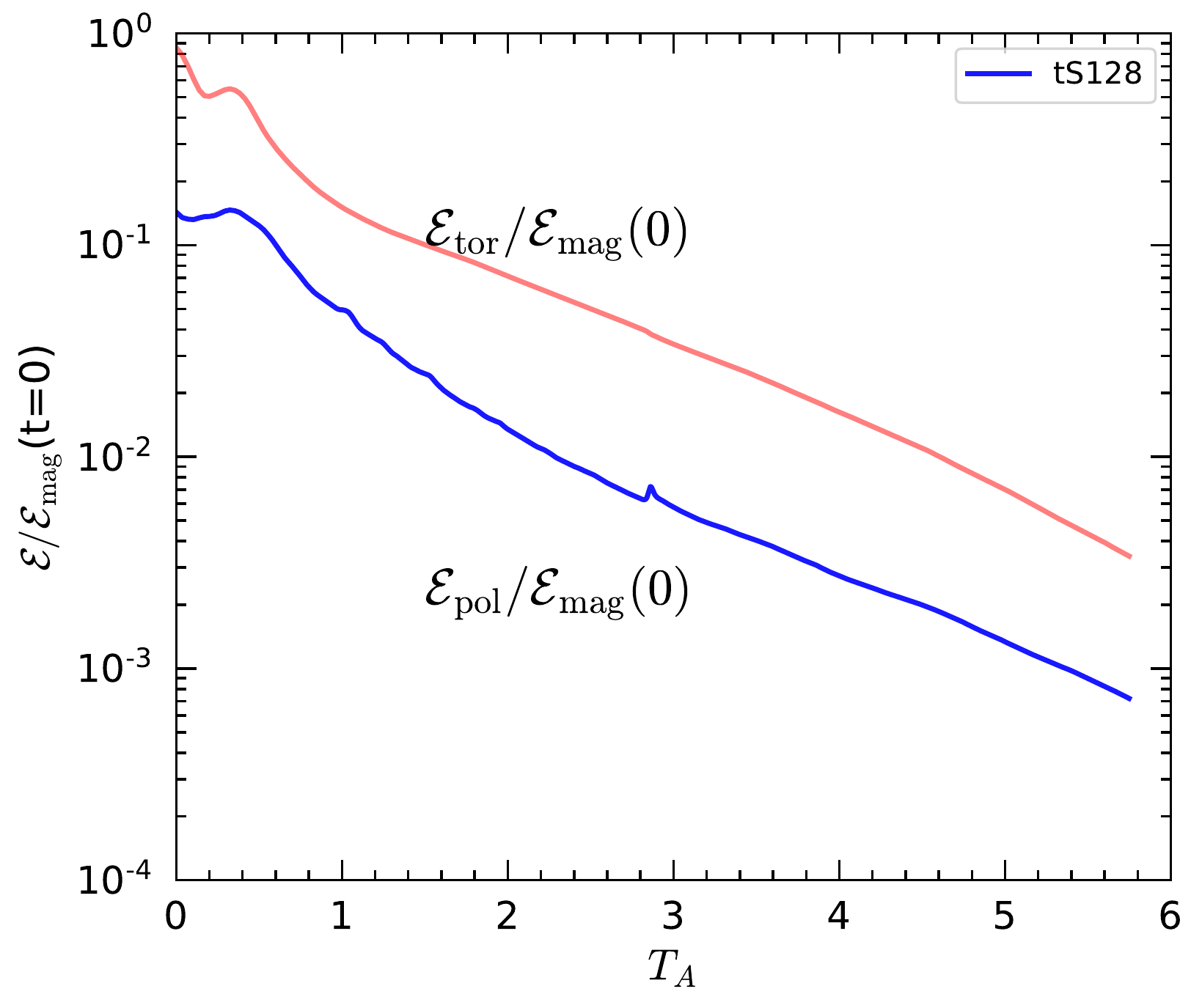}
	\includegraphics[scale=0.35]{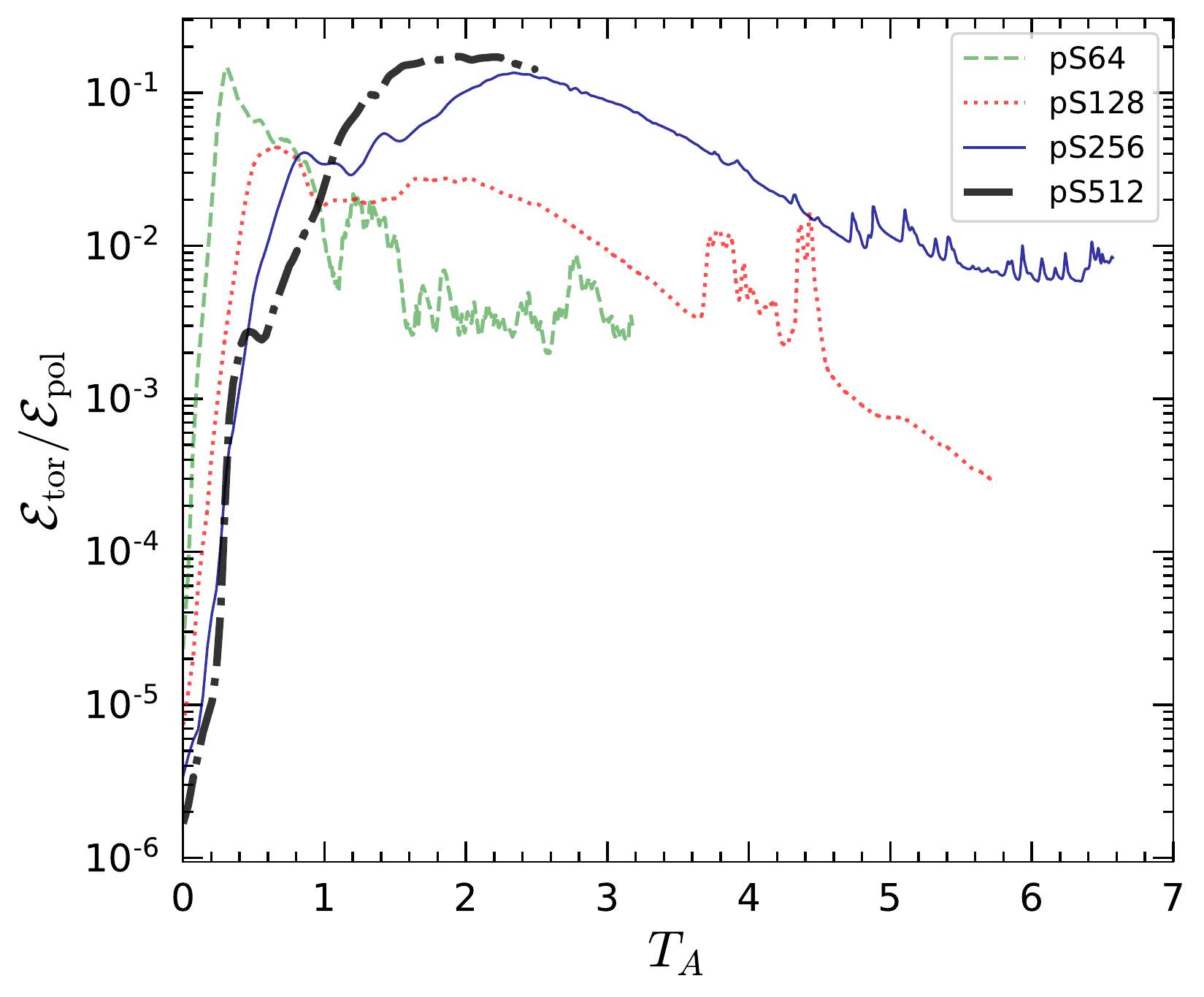}
	\caption{Evolution of poloidal and toroidal energies normalized to the initial magnetic energy (in log scale) for (left) poloidally dominated initial condition (middle) toroidally dominated initial condition, and (right) ratio of toroidal to poloidal energies for the runs \sm{pS}{512}, \sm{pS}{256}, \sm{pS}{128}, and \sm{pS}{64}.}
	\label{energies}
\end{figure*}

\subsection{Magnetic field lines and energies}
We discuss the purely poloidal initial field first. {The \textit{neutral} line corresponds to the region where the magnetic field vanishes inside the star}. Snapshots of the three-dimensional (3d) view of the magnetic field lines are shown in Figure \ref{fig3d} at different times 2.0 ms (left panel), 12 ms (middle panel), and 49 ms (right panel) respectively. We also show two-dimensional projections of the field lines on the x-y plane (equatorial view) and the x-z plane (meridional view) in Figure \ref{flines} with the title representing different time stamps. The color scale gives the strength of the toroidal component in Gauss. During the start of evolution, the field lines first change their cross-sectional area which corresponds to the so called ``varicose'' mode. This is followed by the transverse displacement of the fluid along the neutral line which leads to the development of the ``kink'' instability \citep{Lander2011,Lasky2011,Lasky2012} as shown in Figure \ref{fig3d} (bottom row centre image).

The saturation of this instability modifies the magnetic field as expected : the initial axisymmetry in the system is replaced by a nonaxisymmetric structure. {The toroidal field initially starts appearing on the boundary of the star along the $x=0$ and $y=0$ axes. This is an artefact of the Cartesian grid and the presence of sharp density gradients across the stellar surface at $R=10$ km. Similar features have also been noted in past studies \citep{Lasky2011} and depend strongly on the resolution of our setup, since these toroidal areas become considerably smaller with higher resolution. The toroidal field grows exponentially from the initial state until $t\sim 12 \, {\rm ms} (T_A \sim 1)$ appearing not only within the closed field regions inside the star, but also outside it which might be caused by the aforementioned artefact. The toroidal field strength at this point is comparable to the poloidal field (see middle panel Figure \ref{flines})}. From the equatorial view, we see that the field lines create vortex-like structures (see right panel) owing to the conservation of magnetic helicity. From $t=12$ ms, the evolution proceeds with nonlinear rearrangement of the field, where not only the closed field lines are involved but the whole star and the open field lines. The equatorial 3d view (last panel) shows the already formed toroidal component. The evolution of the field occurs slowly in which the interior closed field lines move outwards losing energy in the toroidal component. We shall see this in more detail when discussing the energies of the poloidal and toroidal components later in this section. The modifications of the field also expels matter from the star however the change in rest mass ($\Delta M_{\rm rest} \approx 10^{-5} \Msun $ ) is much slower as compared to the changes in magnetic field energy during the first $\alfven$  crossing time. The extent to which the field modifies obviously depends on the magnetic field strength with stronger fields having more violent dynamics and vice versa \citep{Ciolfi2012}. From the end state of our simulation, we conclude that the geometry of the magnetic field is significantly different and has left no trace of its initial configuration. Although we do not have resistivity, there is numerical dissipation from our grid and it is difficult to understand whether the post-instability configuration is stable. However, the timescale on which other quantities such as the rest mass of the star are changing is much longer than the instability of the magnetic field. We can conclude that this configuration is not a ``strict'' equilibrium but rather a ``quasi-stationary'' equilibrium.  

The magnetic flux decays in the interior as well as in the exterior of the star due to the rearrangement of the field which is a physical effect and not a numerical artifact as also seen in \cite{BraithwaiteSpruit2005}. We recall that our setup includes an MHD fluid ball with an atmosphere with no solid crust being present. This has important consequences for the magnetic field evolution both inside and outside the star. We do not also have resistive MHD effects such as magnetic reconnection. Including these effects, we would expect powerful outbursts from the atmospheric emission originating from sudden rapid rearrangement of the field. \cite{Thompson1995} proposed that a large-scale reconnection/interchange instability of the magnetic field caused the 1979 March 5 burst event and that cracking of the NS crust produced soft gamma repeaters. Magnetic field decay could also build stresses in the NS crust and cause it to break due to a strong toroidal field, resulting in crustquakes. Magnetar giant flares may likely be explained by such a phenomenon \citep{Lander2015}. 

{Figure \ref{bratio_p} shows the time evolution of the poloidal ($\epol$) and toroidal energies ($\etor$) calculated over the volume of the star defined as:}
\begin{align}
&\mathcal{E}_{pol} =  \int_{\rm star} (b_P^2 u_{\mu}u^{\nu} + \frac{b_P^2}{2}g_{\mu}^{\,\,\,\nu})\delta_t^{\mu}n_{\nu}d^3\vec{x}\sqrt{-g}\\
&\mathcal{E}_{tor} = \int_{\rm star} (b_T^2 u_{\mu}u^{\nu} + \frac{b_T^2}{2}g_{\mu}^{\,\,\,\nu})\delta_t^{\mu}n_{\nu}d^3\vec{x}\sqrt{-g}
\end{align}
where $b_P^{\mu} + b_T^{\mu} = b^{\mu}$ is magnetic field components projected into a space normal to the fluid four velocity $u^{\mu}$, $n_{\mu}$ is the normal observer's four velocity, $g_{\mu \nu}$ is the metric and $\sqrt{-g}=\alpha\sqrt{\gamma}$ is the determinant of the metric tensor. The magnitude of the toroidal field is given by
\begin{eqnarray}
b_T = \sqrt{g_{xx}}\frac{\big(xb^y-yb^x\big)}{\sqrt{x^2+y^2}}
\end{eqnarray}
where $x,y$ are the Cartesian axes.

\begin{figure*}
	\centering
	\includegraphics[scale=0.4]{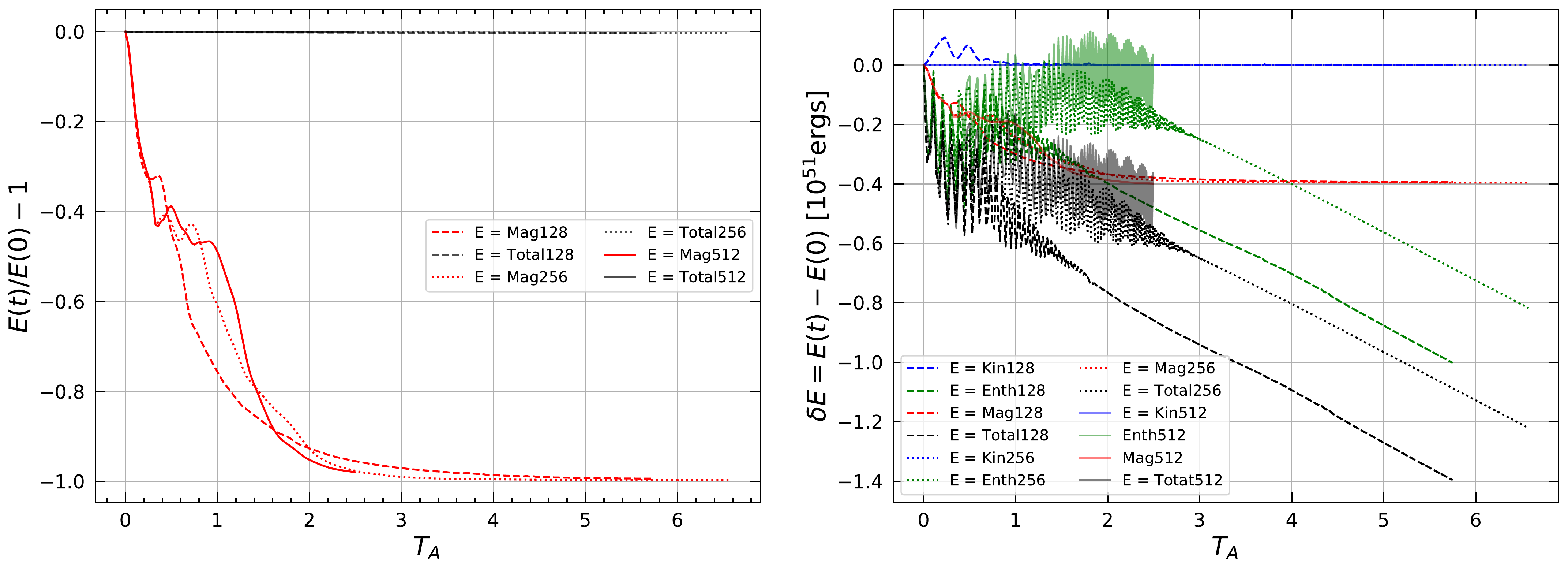}
	\caption{ (Left) Difference in $E(t)$ and $E(t=0)$  normalized by $E(t=0)$ for the total energy (black) and magnetic energy (red). (Right) Difference in kinetic energy (blue), magnetic energy (red), total energy (black) and enthalpy (green) to its initial value for the different setups \sm{pS}{128}, \sm{pS}{256}, \sm{pS}{512} with respect to $T_A$.}
	\label{bkenergy}
\end{figure*}

We compute various energy integrals whose definitions are given in the appendix while the definitions of the different physical variables can be found in \cite{Noble_2006}. Initially at $t=0$, we have the entire magnetic energy stored within the poloidal component. For \sm{pS}{256}, we find the toroidal energy rises peaks at $t=60$ ms ($\sim 2 T_A$) where $\etor\sim 0.2 \emag$ (left panel Figure \ref{bratio_p}). However, as the system loses magnetic energy, the toroidal component gets weaker and at much later times $t\sim 880\,$ ms ($\sim 6.5 T_A$), it approximately becomes 1\% of $\emag$. The evolution shows that the toroidal field attains a quasi-stable equilibrium with energies similar to the ones obtained from solving the Grad-Shafranov equation (see e.g. \cite{Lander2009, Armaza2015, Sur2021}) which gives equilibrium solutions but doesn't say anything about the stability of these equilibrium fields. {The star continues to lose energy till the end of our simulation but the ratio of poloidal and toroidal energies to the total magnetic energy is seen to settle at a quasi-equilibrium value for \sm{pS}{64} and \sm{pS}{256} but not in \sm{pS}{128} (Figure \ref{bratio_p} left panel). In our setup, \sm{pS}{512}, we have a shorter evolution time, $T_A \sim 2.5$, and need a much longer evolution to understand if it reaches equilibrium. The toroidal field in this case grows upto $20 \%$ of the total magnetic energy which is achieved at $2\, T_A$. From this point, although we do not have results for further evolution, its strength seems to decrease and we may fail to achieve convergence as also seen in our different setups in the later stages of evolution}. For \sm{tS}{128} (see right panel of Figure \ref{bratio_p}), we initially have a larger toroidal component which at first loses some energy and increases the strength of the poloidal component. However, this soon becomes stronger with time and the poloidal component also stabilizes at $\epol \approx 0.2 \emag$. Note that this is different than what was observed in \cite{Sur2020} where the toroidal energy decayed to get stabilized at $10\%$ of $\emag$. {This may be caused due to the implementation of the boundary conditions of the magnetic field. In \cite{Sur2020}, we used periodic boundary in the azimuthal direction while the radial and angular boundaries were fixed at their dipolar poloidal values. Since we have free boundary in this setup, the toroidal field gets stabilized by a weaker poloidal component.}

Let us now discuss the dynamics at the initial stage of the evolution. The poloidal field energy remains unchanged up until 12 ms ($\sim 1T_{A}$). During this time, the toroidal field undergoes an exponential growth, and after $\sim 1T_{A}$ the instability saturates and the field continues to evolve less dramatically. This behaviour depends on the initial strength of the magnetic field \citep{Ciolfi2012} and the resolution of our numerical grid, however it is evident that the poloidal energy undergoes this sharp decrease when the instability saturates and the nonlinear rearrangement of the field starts. {The poloidal and toroidal energies (normalized by the initial magnetic energy at $t=0$) shown in the left panel of Figure \ref{energies} are in good agreement with \cite{Ciolfi2012} till the end of their simulation time $t\sim 60$ ms. We have much longer runs in which we see that the component energies continue to decay, both for the poloidally dominated setup (left panel in Figure \ref{energies}) and the toroidally dominated setup (see middle panel of Figure \ref{energies}). The ratio of toroidal to poloidal energies (right panel of Figure \ref{energies}) also shows that for \sm{pS}{128} and \sm{pS}{64}, the field oscillates and dissipates more energy when compared to \sm{pS}{256} where the energy loss is more continuous. {This ratio for the setup \sm{pS}{512} reaches 20\% at $\sim 2T_A$ while the setup \sm{pS}{256} reaches 10\% at $\sim 2.5 T_A$. Again, we lack convergence in our simulations, however, at later stages at $6.5 T_A$, the toroidal field for \sm{pS}{256} settles at 1\% of the poloidal energy.}

\begin{figure*}
	\includegraphics[scale=0.47]{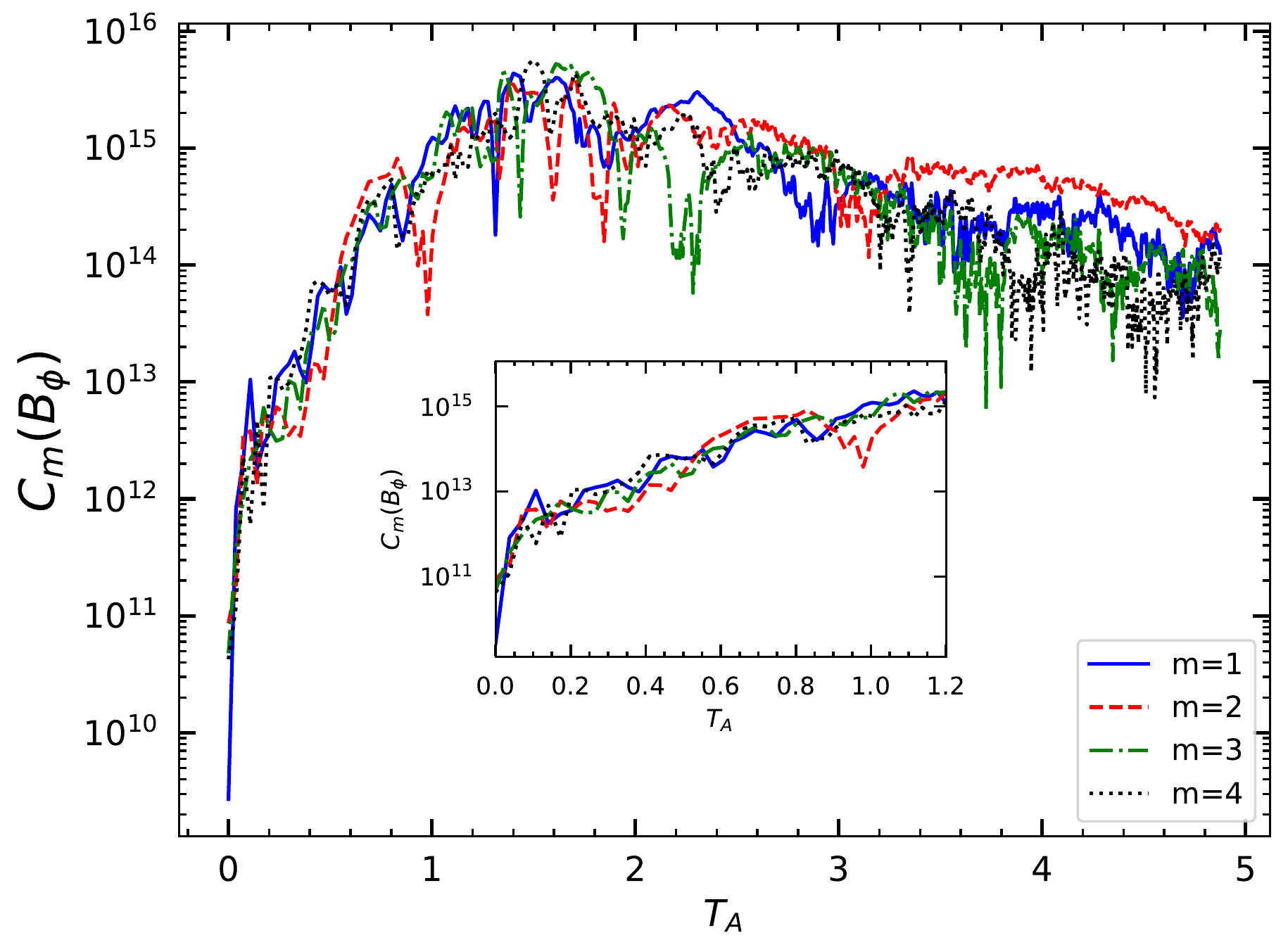}
	\includegraphics[scale=0.47]{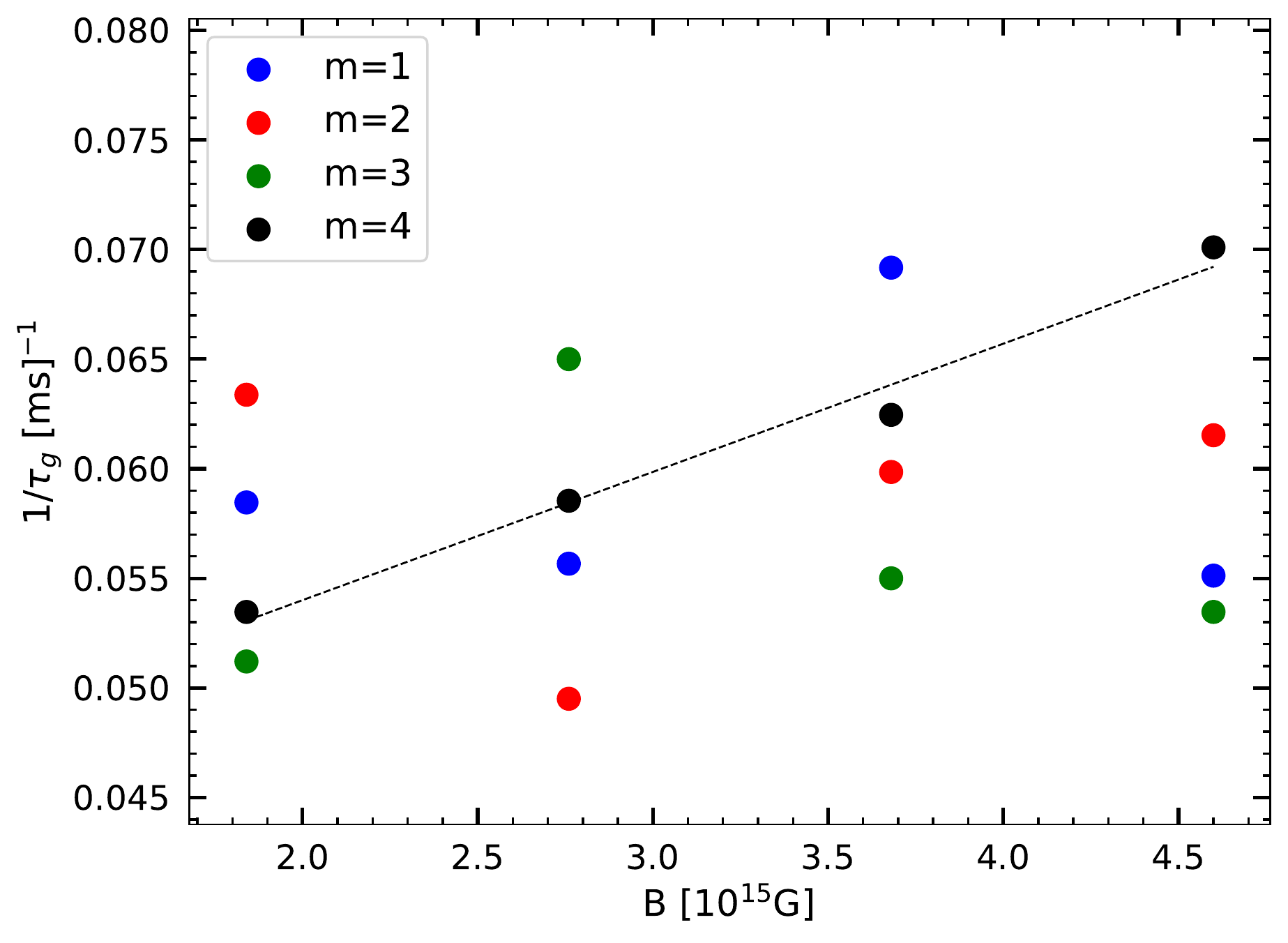}
	\caption{(Left) Fourier decomposition of $B_{\phi}$ {for the setup \texttt{pS256}} and complex weighted averages for the different modes $m\in(1,2,3,4)$ shown till 5 $T_A$. The inset shows the first 1.2 $T_A$ of the evolution where we can see that the $m=2$ grows the fastest till $1 \, \alfven$ crossing time. (Right) Linear relationship between the $\alfven$ times calculated for the different magnetic field strengths in \sm{pS}{128}.}
	\label{cm}
\end{figure*}

To study the enrgetics, we decompose the total energy of the star into four different components following \cite{Noble_2006}: kinetic, magnetic, rest mass and the enthalpy (see the Appendix of this paper for the mathematical expressions). In the left panel of Figure \ref{bkenergy}, we show the change in magnetic and total energies to its initial value ($\delta E$) normalized by the initial value. We see that the total energy remains conserved, however the magnetic energy decays. To understand the loss of magnetic energy, we plot $\delta E$ for the individual energy components (except the rest mass as it remains conserved in our simulations) inside the star for {the setups \sm{pS}{128},\sm{pS}{256}, and \sm{pS}{512},} and look at their behavior with $T_A$. First, the change in kinetic energy from its initial value is negligibly small which means that the fluid almost remains static and experiences only small variations in movement due to the presence of the instability which we notice at $\sim 0.5 T_A$. Second, the magnetic energy decreases and this loss is independent of the resolution of our simulations. More than $90\%$ of the initial magnetic energy is lost which either gets radiated away to infinity or gets dissipated as heat inside the star. We calculated the Poynting flux over the surface of the star and found that ${\sim}6.84\times10^{44}$ ergs of the total magnetic energy gets converted to radiation outside. However, a major portion goes into heating the interior of the star as we can see from the rise in enthalpy till ${\sim} 2T_A$. The enthalpy loss is dependent on the grid resolution and reduces by a factor two when the resolution is increased of the same factor. Higher resolutions than those considered here would be needed to minimize numerical dissipation effects. {These results strongly depend on the outflow boundary conditions used in our simulations; however, realistic NS has a crystalline-solid crust, which would prevent any dissipation of magnetic energy outside the star and can significantly influence our simulation results.}

\subsection{Growth}

Following \cite{Zink2007,Lasky2012}, we study the magnetic field dynamics in terms of the Fourier modes
\begin{equation}
C_m = \int_{0}^{2\pi} B_{\phi}(\bar{\omega},\phi,z=0)e^{im\phi}d\phi\,,
\end{equation}
where $\bar{\omega} = \sqrt{x^2+y^2} = 0.8R$ is a contour in the equatorial plane of the star. We compute $C_m$ for $m \in [1,2,3,4]$ and show them in Figure \ref{cm}.

{The inset gives a closer look at the initial stage of the evolution. As a first observation, we can see that all the different modes are excited and each one grows exponentially. Secondly, this growth saturates after few $\alfven$ periods from which the different modes evolve less dramatically. Thirdly, the loss of magnetic energy causes these modes to lose strength. And lastly, all the different modes grow closely (as seen from the inset of figure \ref{cm} left), however, the $m=2$ mode remains strongest followed by $m=1$, $m=4$ and $m=3$ respectively. We calculate the growth times ($\tau_g$) for the various modes defined by the following:}

\begin{eqnarray}
\tau_{g} = \frac{\Delta t}{\Delta {\ln}(C_m)}
\end{eqnarray}

{The strength of the volume averaged magnetic field is varied in our simulations using the setup \sm{pS}{128} since it is computationally less expensive to run each simulation than \sm{pS}{256}}. By using the modes $(1,2,3,4)$, we calculated the growth times for each case.} We plot the inverse of $\tau_g$ as a function of $B$ in Figure \ref{cm} (right). There are some errors introduced when computing $\tau_{g}$ during the exponential phase as it is difficult to select an interval of time where this growth happens and one should take different realizations and report the mean and standard deviation of these data. Since we are only qualitatively interested in the behavior of $\tau_{g}$, we take the time interval between the minimum and maximum values of $C_m$. {It is difficult to establish a linear relationship between $\tau_g$ and the inverse field strength predicted by perturbation theory, but we found one only for the mode $m=4$ and presented the best fit dashed line in figure \ref{cm} (right). The values of $\tau_{g}$ for the field strengths $\{4.5,3.6,2.7,1.8\}\times10^{15}$ G are respectively the following: $\{18.1,14.4,17.9,17.1\}$ ms for the mode $m=1$; $\{16.2,16.7,20.2,15.7\}$ ms for the mode $m=2$; $\{18.7,18.1, 15.4,19.5\}$ ms for the mode $m=3$; and $\{14.2,16.0,17.0,18.7\}$ ms for the mode $m=4$. The growth time for the strongest magnetic field setup should be shortest for all modes considered, but this is observed only for $m=3$ and $m=4$. A majority of these modes have abrupt erratic growth times, with the shortest mode being $m=3$ for the field strength of $3.6\times10^{15}$ G while the longest mode was $m=2$ for the field strength $2.7\times10^{15}$ G.}
\begin{figure*}
	\includegraphics[scale=0.35]{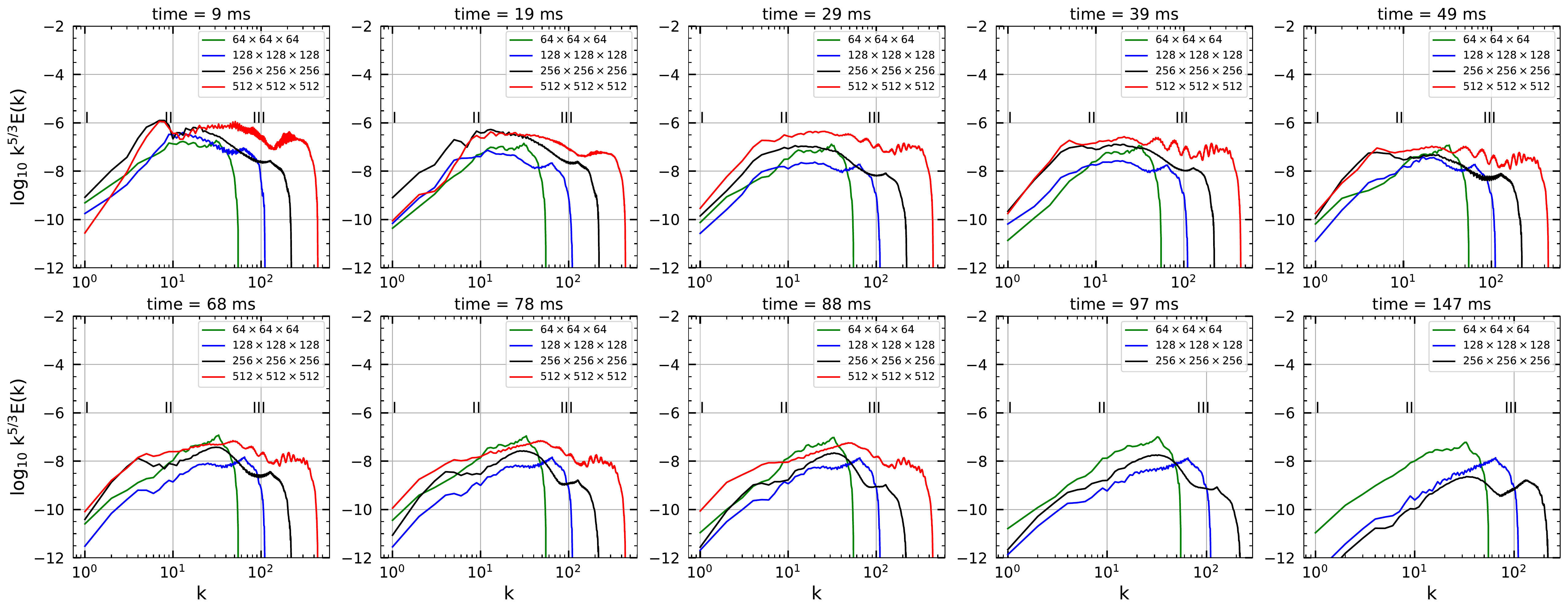}
	\caption{{Spectra of specific kinetic energy $E(k)$ (equation \ref{ek}) inside the star and multiplied by Kolmogorov scaling $k^{5/3}$} calculated for our different resolution setups (given as figure labels). The different plots represent different times in the evolution of our system (given by the title in each figure).}
	\label{spectra}
\end{figure*}

\begin{figure*}
	\includegraphics[scale=0.35]{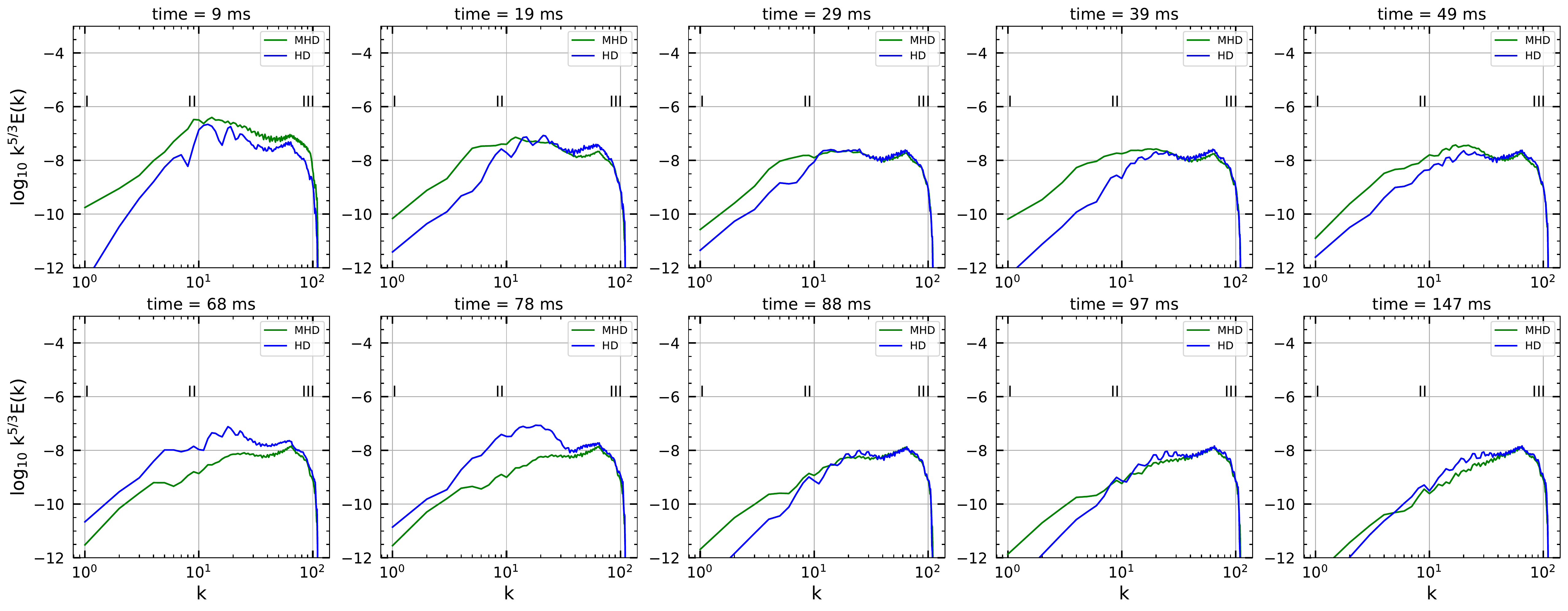}
	\caption{A comparison of the specific kinetic energy spectra for \sm{pS}{128} between the HD simulation and the MHD simulation. }
	\label{spectra3}
\end{figure*}

\subsection{Turbulence}

{The magnetic field in young NSs experiences turbulence after their birth during a period when the neutrinos have not fully escaped and provide heat as energy \footnote{In private communication with Andreas Reisenegger.}  \citep{Mabanta2018}. In old NSs, curvature contributions to mutual friction also cause differences in angular velocity between the superfluid and the rest of the star, which can also cause superfluid turbulence in the core of the NS \citep{Peralta2006,Andersson2007}.} It is expected that the magnetic field should break in smaller eddies and dissipate most of its energy as the star ages, but observations show that a certain fraction of NSs population has stronger fields. Longer timescale ($\sim 1$ Myr) simulations have shown that the magnetic field in the crust attends a state known as ``Hall-attractor'' after driven by the Hall effect \citep{KGAC2014}. Even though this equilibrium stage is achieved only at much later times, the magnetic field evolves more rapidly in the initial phase in the lives of NSs. {As a result of the initial instability and reconnection, turbulence can be driven by the magnetic field in NSs.}

On simulating magnetic field evolution in NSs, it was observed that the system exhibited turbulence when the Kolmogorov spectra were calculated for both the kinetic and magnetic energies \citep{Sur2020}. This was speculated to be caused by the initial perturbation of the fluid velocity which was used to trigger the instability quickly. However, the signals obtained were noisy, owing to the limited resolution of the simulations, {and this caused the spectra to be indistinguishable for the power-law scaling of $-5/3$ (predicted by the classical Kolmogorov theory \citep{Kolmogorov1941}) from other MHD turbulence spectra \citep{boldyrev2011,beresnyak2019,Schekochihin2020}. The fact that the poloidal and toroidal energies in the simulations presented in \cite{Sur2020} did not settle to an equilibrium, but rather oscillated around a mean value, could possibly be the result of turbulence triggered by the instability of the initial purely poloidal field. To understand whether this turbulence is physical and not caused by numerical effects,
it is required to analyse simulation data generated from higher resolution and longer MHD runs.}

In this work, we are in a better position to study the effect of turbulence given that our simulations have higher resolution compared to ones presented in \cite{Sur2020}. We compute the power spectrum of the specific kinetic energy. The Fourier transform of the three-dimensional velocity vector is given by
\begin{equation}
{u^{j}}(\mathbf{k}) = \int {u}^{j}(\mathbf{x})e^{-i\pi\mathbf{k}\cdot\mathbf{x}}d^3\vec{x}
\end{equation}
and the specific kinetic energy is calculated as
\begin{equation}
\varepsilon_k (\mathbf{k}) = \frac{1}{2}u^{j}(\mathbf{k})\cdot{u^{j}}^{\dag}(\bf{k})
\end{equation}
where ${u^{j}}^{\dag}$ is the complex conjugate of ${u^{j}}$. It is then straight forward to calculate the velocity power spectral density (PSD) as
\begin{equation}
E(k) = \frac{1}{2}\Sigma_{{\bf{k}<|\bf{k^{\prime}}|<k+dk}} \varepsilon_k(\bf{k^{\prime}})
\label{ek}
\end{equation}
We use the numerical algorithm outlined in \cite{Navah2018}. We select our computational domain such that it filters out regions only within the stellar surface with a radius of $0.95 R$. For this, we use the density value $\rho\geq10^{14}$ {$\gccm$} to select the sphere and discard regions with $\rho<10^{14}$ {$\gccm$}. As the strong magnetic field deforms the star, this selection criteria correctly accounts the effect of non-sphericity. Due to the sharp gradients in velocities present close to the surface, we apply an exponential decay window function and make the velocities go to zero when $r$ approaches $R$. {The window function is only applied in post-processing, i.e. while computing the turbulent spectra, and not during the evolution of the magnetic field in our simulations. This windowing is required to make the data segment periodic and prevent any jump discontinuities that may arise while computing the Fourier transform of our velocities. Without the window function, there arises superficial features in the spectra that are unphysical.} Figure \ref{spectra} shows the specific kinetic energy spectra vs wavenumber ($k = |\bf{k}|$) for the different resolutions setups with varying times. {The red, black, green, and blue lines corresponds to the setups \sm{pS}{512}, \sm{pS}{256}, \sm{pS}{128} and \sm{pS}{64} respectively. Since the setup \sm{pS}{512} has an evolution time of 88 ms, we show its spectra until this point. For convenience, we plot the quantity $k^{5/3}E(k)$ to get a flat spectrum when it follows the Kolmogorov scaling. There are three distinct regions characteristic of such a spectrum: the energy-containing range (region I), the inertial subrange (region II), and the dissipative range (region III) as shown in the figures. At $t=29$ ms and $t=39$ ms, we see that the spectra in region II becomes flat but the Kolmogorov scaling becomes weaker as seen from $t=68$ ms. The evolution shows us that at largest scales, kinetic energy is gradually lost, while this is not so prominent for our lower resolution setup. The spectra at different times indicate that there is a lack of convergence, as we should expect that at large scales, the spectra should have similar energies regardless of the resolution.}

To understand whether this feature is physical, we performed a pure hydrodynamics (HD) simulation for \sm{pS}{128} without evolving the magnetic field. We calculated the spectra for the pure hydro run and compared with the MHD simulation in figure \ref{spectra3}. The spectra differed at large scales where MHD is seen to drive large scale flows while the dynamics for the hydro run at small scale comparatively remains same at different times caused by atmospheric noise. This was also visible when we plotted the velocity field of our star and observed large spikes at y=0 axis on the equatorial plane. {Thus, most likely at small scales, we are not observing turbulence, but rather a noisy velocity field in our simulations.}

%% file: conclusions.tex
In this paper, we presented {long-term} GRMHD evolution in NSs by performing simulations using the code {\tt Athena++}. We studied the energy variations of the poloidal and toroidal magnetic fields, the kinetic energy and the enthalpy with time for {four different resolutions $64^3$, $128^3$, $256^3$ and $512^3$ in Cartesian grid.}

We explored two different initial conditions, one purely poloidal and one with dominant toroidal field. We find that in all the different resolution setups, a purely poloidal field is unstable and this gives rise to a toroidal component. The toroidal energy becomes comparable in strength to the poloidal energy during the initial stages of the evolution, but at later times, it decreases significantly and becomes approximately $1\%$ of the total magnetic energy at $t\sim 880$ ms corresponding to $6.5 \, \alfven$ periods. {Our setup $\sm{pS}{512}$ has an evolution time of $2.5 T_A$ at which the toroidal field reaches 20\% of the poloidal energy and $10\%$ of the total magnetic energy once the simulation ends. However, our longer simulations such as \sm{ps}{256} do not reach any equilibrium magnetic field configuration and lack convergence at later stages in the evolution.} {For the toroidally dominated setup, we found the ratios of poloidal and toroidal energies to the total magnetic energy to settle at an equilibrium value of 0.2 and 0.8 respectively. }

On comparing the different initial conditions, we address two main issues. The first concerns the different behaviour of the toroidal vs poloidal dominated simulations. The  stronger toroidal setup develops a sizable poloidal component with 20\% of the total magnetic energy but does not become the dominant component at the end of the simulation. On the other hand, the purely poloidal setup does not develop such a large toroidal component which is likely due to the different boundary conditions implemented in this work when compared to \cite{Sur2020}. The second issue, concerns the final values of the different energy components. Our simulations are significantly longer than those in \cite{Sur2020}. 

In Fig.~\ref{energies} it appears that the ratio between toroidal and poloidal energies, in the case of our higher resolution simulations, is gradually decreasing. Moreover, the higher resolution simulations seem to be losing more magnetic energy compared to the lower ones. This loss in magnetic energy from the star increases the internal energy, while around $10^{44}$ ergs of the magnetic energy are radiated to infinity in the form of electromagnetic radiation. {Our models, however, do not have a solid crust or resistivity, two important factors that influence electromagnetic emission in realistic NSs. Some of the most luminous magnetar giant flares can be explained with the release of energy from crustal breaking \citep{Lander2015}.}

An important aspect we addressed in our simulations was to study the turbulence in NS MHD simulations. It was seen in \cite{Sur2020} that the magnetic field instability caused the system to experience turbulence and this may have caused the poloidal and toroidal energies to reach only a quasi-equilibrium. However, due to limited resolution, the spectra were noisy and difficult to establish a power-law scaling according to the Kolmogorov theory. In this work, we analyzed data from higher resolution MHD runs and found that this turbulence is not physical but rather caused by noise in the velocity field inside the star. This was confirmed on comparing the spectra between HD and MHD runs. The main difference occurred at large scales where the MHD {simulations demonstrated} large scale flows while the small scale dynamics remained the same between HD and MHD cases.

Overall, we find consistent results with the previous GR {works at early times (e.g. \cite{Ciolfi2011, Ciolfi2012} and references therein) {while extending the simulations to 880 ms which is much longer than previously obtained (for example 400 ms in \cite{Lasky2011}}. Our finest grid resolution is $0.1155$ km which is similar to \cite{Ciolfi2012} but also higher than \cite{Lasky2011} where a grid of $\sim 0.23$ km was used.  However, higher resolution and longer simulations} are still required to settle the issue of what happens to the {late time evolution of} different energy components and the turbulence in studies of magnetic field simulations of NSs.

%% file: appendix.tex
Let us first note the following definitions, where the normal observer's four velocity ($n_{\mu}$) in the coordinate basis {$t,x_1,x_2,x_3$} is given by
\begin{eqnarray}
n_{\mu} = (-\alpha,0,0,0)\\
\alpha^2 = -1/g^{tt} = -g_{tt}\\
\gamma = -n_{\mu}u^{\mu} 
\end{eqnarray}
where $\gamma$ is the {Lorentz} boost and $\alpha$ is the known as the lapse function. The projection tensors, which projects into a space normal to the fluid four velocity, are given by:
\begin{eqnarray}
h_{\mu\nu} = g_{\mu\nu} + u_{\mu}u_{\nu}\\
j_{\mu\nu} = g_{\mu\nu} + n_{\mu}n_{\nu}
\end{eqnarray}
We work in isotropic coordinates such that $g_{xx}=g_{yy}=g_{zz}$ and $g_{ij}=0$ for $i\neq j$. The fluid is described by four velocity $u^{\mu}$, rest mass density $\rho_0$, and pressure p. The magnetic field four vector is given by
\begin{equation}
\mathcal{B}^{\mu} \equiv -n_{\nu}F^{\mu\nu}
\end{equation} 
such that $b^{\mu} = h^{\mu}_{\nu}\mathcal{B}^{\nu}/\gamma$. The stress-energy tensor can be decomposed into a fluid part (subscript F) and a magnetic part (subscript B) as following:
\begin{eqnarray}
T^{\mu\nu} = T^{\mu\nu}_{F} + T^{\mu\nu}_{B}
\end{eqnarray}
where
\begin{eqnarray}
T^{\mu\nu}_{F} = w u^{\mu}u^{\nu} + pg^{\mu\nu} \\
T^{\mu\nu}_{B} = b^2u^{\mu}u^{\nu} + b^2g^{\mu\nu} - b^{\mu}b^{\nu}
\end{eqnarray}
where $w=p+\rho_0+\rho_0\epsilon$ and $\epsilon = p/  (\Gamma-1)\rho_0$, is the specific internal energy density. We define the fluid and magnetic currents as
\begin{eqnarray}
J^{\nu}_{F} = -T^{\mu\nu}_{F}\partial{^t_{\mu}}\\
J^{\nu}_{B} = -T^{\mu\nu}_{B}\partial{^t_{\mu}}\\
\end{eqnarray}
such that the total energy current is $J^{\nu}_{E} = J^{\nu}_{F} + J^{\nu}_{B}$ and the total energy is
\begin{eqnarray}
E = \int (-J^{\nu}_E n_{\nu})\sqrt{-g}d^3\vec{x}
\end{eqnarray}
Using the relation $g_{\mu\nu}u^{\mu}u^{\nu}=-1$, we get $u^{t} = \alpha\sqrt{1+g_{xx}(u^1)^2+g_{yy}(u^2)^2+ g_{zz}(u^3)^2}$. The different energies can be computed as following:

\begin{align}
&E = \int T_{\mu}^{\,\,\,\nu}\partial_t^{\mu}n_{\nu}d^3\vec{x}\sqrt{-g}\\
&M_{\rm rest} = -\int \rho_0 u^{\mu}n_{\mu}d^3\vec{x}\sqrt{-g}\\
&E_{B} = \int (b^2 u_{\mu}u^{\nu} + \frac{b^2}{2}g_{\mu}^{\,\,\,\nu}-b_{\mu}b^{\nu})\partial_t^{\mu}n_{\nu}d^3\vec{x}\sqrt{-g}\\
&E_{k} = \int (w u_{\mu}u^{\nu}+p g_{\mu}^{\, \,\, \nu})\partial_t^{\mu}n_{\nu}d^3\vec{x}\sqrt{-g}-M_{\rm rest}-E_{H}\\
&E_{H} = \int (\rho_0\epsilon_{H}+p)\gamma d^3\vec{x}\sqrt{-g}
\end{align}
where $\sqrt{-g} = g_{xx}^{3/2}$, E is the total energy, $E_B$, $E_k$ and $E_H$ are the magnetic, kinetic and enthalpy respectively. 